\documentclass[journal]{IEEEtran} 
\IEEEoverridecommandlockouts

\usepackage{cite}
\usepackage{amsmath,amssymb,amsfonts}
\usepackage{algorithmic}
\usepackage{graphicx}
\usepackage{textcomp}
\usepackage{xcolor}
\usepackage{gensymb}
\usepackage[utf8]{inputenc}
\usepackage{mwe}
\usepackage{float}
\usepackage[caption = false]{subfig}
\usepackage{enumitem}
\usepackage{changepage}
\usepackage[flushleft]{threeparttable}
\usepackage{booktabs,ragged2e}
\usepackage{multirow}
\usepackage{url}
\usepackage[linesnumbered,ruled,vlined]{algorithm2e}
\SetKw{Continue}{continue}
\SetKw{Not}{not}
\usepackage[symbol]{footmisc}
\usepackage{hyperref}
\usepackage{pdfpages}

\begin{document}

\title{\emph{PUF-Phenotype}: A Robust and Noise-Resilient Approach to Aid Intra-Group-based Authentication with DRAM-PUFs Using Machine Learning}

\author{	
%	\vspace{0.2cm}
 Owen Millwood, Jack Miskelly, Bohao Yang, Prosanta Gope \textit{Senior Member, IEEE}, Elif Bilge Kavun \textit{Member, IEEE}, and Chenghua Lin 
%\IEEEcompsocitemizethanks{\IEEEcompsocthanksitem O. Millwood, B. Yang, P. Gope, and C. Lin  are with Department of Computer Science, University of Sheffield, Regent Court, Sheffield S1 4DP, United Kingdom.
%J. Miskelly is with the Centre for Secure Information Technologies, Queen's University Belfast.
%E. B. Kavun is with the Secure Intelligent Systems Research Group, FIM, University of Passau.
%%% note need leading \protect in front of \\ to get a newline within \thanks as
%%% \\ is fragile and will error, could use \hfil\break instead.
%(E-mail: prosanta.nitdgp@gmail.com/p.gope@sheffield.ac.uk)
%(E-mail: elif.kavun@uni-passau.de)
%\IEEEcompsocthanksitem

%\textbf{Corresponding author:} Dr. Prosanta Gope
%%Corresponding author: Prosanta Gope
}
%\IEEEcompsocthanksitem
%}% <-this % stops a space}
\maketitle

\begin{abstract}
  As the demand for highly secure and dependable lightweight systems increases in the modern world, Physically Unclonable Functions (PUFs) continue to promise a lightweight alternative to high-cost encryption techniques and secure key storage. While the security features promised by PUFs are highly attractive for secure system designers, they have been shown to be vulnerable to various sophisticated attacks - most notably Machine Learning (ML) based modelling attacks (ML-MA) which attempt to digitally clone the PUF behaviour and thus undermine their security. 
  More recent ML-MA have even exploited publicly known helper data required for PUF error correction in order to predict PUF responses without requiring knowledge of response data. In response to this, research is beginning to emerge regarding the authentication of PUF devices with the assistance of ML as opposed to traditional PUF techniques of storage and comparison of pre-known Challenge-Response pairs (CRPs). In this article, we propose a classification system using ML based on a novel \textit{`PUF-Phenotype'} concept to accurately identify the origin and determine the validity of noisy memory derived (DRAM) PUF responses as an alternative to helper data-reliant denoising techniques. To our best knowledge, we are the \emph{first} to perform classification over multiple devices per model to enable a group-based PUF authentication scheme. We achieve up to 98\% classification accuracy using a modified deep convolutional neural network (CNN) for feature extraction in conjunction with several well-established classifiers. We also experimentally verified the performance of our model on a Raspberry Pi device to determine the suitability of deploying our proposed model in a resource-constrained environment.
\end{abstract}

\section{Introduction}

IoT devices are an unavoidable and growing presence in modern life and have been adapted for a wide variety of applications, ranging from domestic use to healthcare, autonomous vehicles, and even military systems \cite{iotreview}. Many of these applications have high security and privacy requirements due to the sensitivity of the information they either transmit or store. Providing adequate security to such systems is more challenging due to heavy restrictions in the availability of computational, storage and network resources, making traditional security methods often inaccessible due to their high complexity. As a result, there is a requirement for robust, lightweight solutions to address security concerns. Physically Unclonable Functions (PUFs) have been proposed as one such solution, providing systems with lightweight, reliable, and highly-unpredictable fingerprints to enable strong security at a low resource cost. This security is achieved by deriving a hardware-rooted identity from low-level component variation introduced as an unavoidable byproduct of the manufacturing process. PUF identities can be used for memory-less secure key generation or directly to generate single-use authentication tokens for device verification in a challenge-response protocol. As entropy is extracted from small physical variations in the circuitry, however, environmental effects such as temperature, voltage variation, and temporal effects (ageing) can introduce noise in the PUF response. To overcome this, a degree of error correction is required in most PUF designs. A common approach is to use Helper Data Algorithms (HDAs), which map PUF responses to code words of an Error Correction Code (ECC) scheme. During this process, Helper Data (HD) is required, which must be assumed to be publicly known. Using the HD alongside a given HDA enables useful data to be extracted from the noisy PUF data, such that an expected value is retrieved.

\subsection{Related Work and Motivation} While error correction methods have been shown to be effective for reducing noise from measured PUF responses, in some implementations this can lead to the introduction of new vulnerabilities which target the error correction mechanisms themselves. Recently a new problem has emerged from this issue, namely \textit{privacy leakage through HD} \cite{ML_helper_data}. In a standard PUF threat model, it is assumed that challenges and HD are public, therefore accessible to an attacker. A HD attack is mounted by an attacker who only requires access to challenge data and HD, and allows them to predict response data and thus compromise the PUF secret. This is in contrast to more well-studied modeling attacks, where knowledge of the input challenge and matching PUF response is required for some large number of challenges in order to comprise other unseen response secrets \cite{CCS:RSSDDS10, DelvauxJeroen2019MAoP}. PUF authentication schemes are often accompanied by an extensive security analysis against a common PUF adversary model (which typically includes modeling and/or side channel attacks), however these schemes often also integrate a requirement for HDA and/or ECC to deal with noisy PUF responses \cite{GuChongyan2021AMAR, gope_scalable, KustersLieneke2019SCRf}. An issue with such systems is that they do not regard this new type of HD attacker, who only requires already publicly available data in order to successfully model the PUF. It is therefore essential for techniques to be developed that are able to counter such an attacker, while simultaneously ensuring security and resource requirements are met. While it is common practice for PUF security systems to employ HDA and ECC, it is not a strict necessity, rather an established industry-accepted approach for denoising PUF responses. To achieve sufficient noise reduction by a different means, it is intuitive to look to fields of study where noise reduction is a key focus, such as computer vision. In the field of computer vision, issues of denoising and classification of noisy data are a well-studied topic. Machine Learning (ML) is a powerful tool used both for extracting noise from images, such as feature extraction through Deep Denoising Autoencoders \cite{LiJunhua2015Flfi} and image classification of already noisy images \cite{Roy2018}. The problems solved through these techniques in computer vision are almost directly represented by the noisy PUF problem, namely the requirement to either actively remove noise from PUF measurements or classify the PUF measurement regardless of the noise. One type of PUF where this is particularly relevant are DRAM based PUFs which produce very large amounts of PUF data but with a high degree of noise. With large response sizes, it is intuitive to represent responses as images to test the effectiveness of computer vision techniques to solve such problems. In \cite{deeppuf}, Deep Convolutional Neural Networks (CNNs) are utilised to perform classification of noisy DRAM-PUF responses using image data of the PUF measurements. Their solution uses a custom memory controller to precisely measure PUF responses and map the physical bit-cell layout directly to a grayscale image which \emph{directly represents} the real DRAM properties and structure. In this approach a very large dataset of DRAM-PUF responses are used in training, as input patterns are based on many arbitrary bit string inputs to the memory. In practice, such responses are not suitable for security purposes as with the DRAM-PUF, there are only three types of unique input value that exhibit suitable entropy (this is discussed in detail in Section \ref{data_dependence}).  Yue et al. proposed a similar approach, using very large raw responses from DRAM start-up values to authenticate PUFs without an explicitly stored CRP database \cite{yueDRAM_CNN}. In this approach, various CNNs are tested to classify three DRAM PUFs and extract features to be compared with features known by a trusted entity.
These works have focused on the learnability of raw memory-based PUF data, where CNNs extract features from image data that directly represent the physical organisation of the bit-cell matrices on chip. Achieving accurate image representations of measuring responses from DRAM however is a difficult task, especially during run-time of a device. In addition, potentially useful information could be stored unknowingly in such images - for example charge leakage paths for capacitive cells - which could in theory provide an attacker with information with which to predict cell-failure behaviour of the PUF.\\
It is important to note that one overlooked advantage of ML-based authentication is the removal of CRP database storage requirement, which has some useful implications. As an attacker can read CRPs, it must be assumed in traditional PUF protocols, the CRP database is stored on a system in a trusted environment, limiting device-to-device authentication as devices must be assumed to be publicly accessible. By offloading the task of response scrutiny to an ML model, storing responses on the prover system for comparison with received responses is no longer required. This has the potential to expand current limitations dictated by non-ML PUF-based authentication schemes. This property also provides a potentially significant advantage over current symmetric and public key-based schemes for IoT, where key data must be stored on devices.
\subsubsection*{\textbf{Motivation}}
\begin{figure}[t]
    \centering
    \makebox[0pt]{\includegraphics[width=70mm]{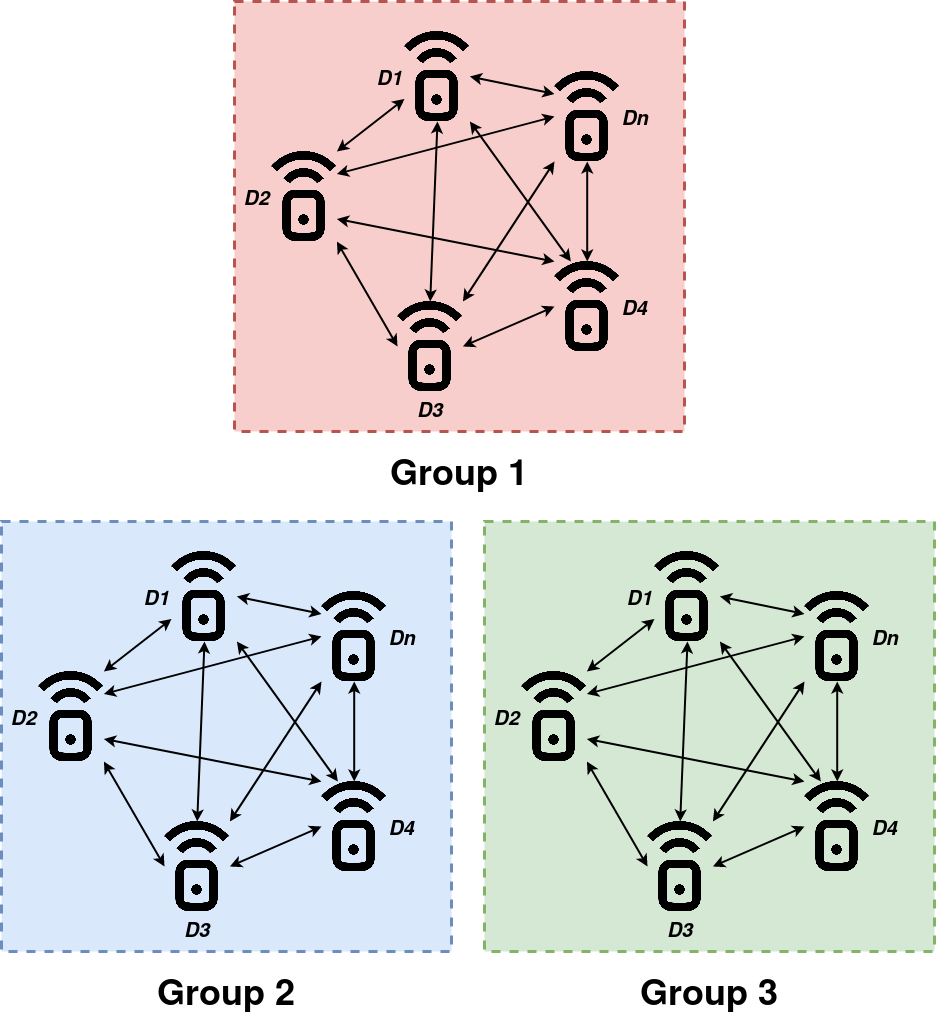}}
    \caption{Intra-Group Communication Environment}
    \label{fig:group_auth}
\end{figure}

While the previously mentioned schemes are proposed as an alternative to error correction algorithms, this is not without trade-off. It must be noted that while the HDA and HD computation and storage requirement is removed, a new computation and storage requirement is introduced. The extremely expensive training of such ML models is performed during enrollment, however, on-device verification still requires model storage and execution. If a device is required to have relatively high computational performance to be compatible with such schemes, the question stands whether a PUF-based solution is suitable over traditional handshaking protocols using asymmetric/symmetric cryptography? Furthermore, the previously mentioned schemes do not fully exploit the benefit of removed CRP database requirement. Each scheme proposes single-authentication with an entity in a trusted environment. These aspects reduce the application of such systems to more traditional device-to-server authentication as opposed to device-to-device. This restriction limits the PUF-based authentication from wider IoT scenarios, for example where a group of devices require inter-node communications within the group without explicit contact with an entity in a trusted environment (Figure \ref{fig:group_auth}). Finally, current ML-based PUF authentication only considers a single device per trained model, therefore considering a group of $n$ provers, a given verifier is required to store $n-1$ models in order to handle authentication requests from each other group member. This is an unrealistic requirement for on-device authentication given the resource constraints of an IoT device. Each of these points raise the question: is it possible to apply a \emph{single} ML-based PUF authentication model that is appropriate for multiple grouped devices, that can run on lightweight devices and without a third party trusted verifier?

\subsection{Contribution}This work proposes a method for authenticating a group of PUF enabled devices using a single combined modified CNN and standard ML classifier that is more lightweight than existing schemes. Our scheme operates on images formed directly from noisy PUF responses without helper data, knowledge of the physical properties of the PUF, or its design. This allows for the use of a single classifier with low storage requirements, which can run on relatively low resource devices, and which can natively perform both group-based authentication (is device X part of group Y?) and device-specific authentication (is device X itself authentic?).
The proposed classification architecture also does not use data which is physically representative of the internal PUF structure but rather an image derived from a non-design specific response data stream. As it relies only on externally observable characteristics and not knowledge of the underlying structures, treating the PUF itself as a black box, we call this approach a \textit{PUF Phenotype} which will be introduced in more depth in Section \ref{phenotype_section}. Finally, unlike existing schemes, our classification architecture is only trained using the three types of input pattern to DRAM which exhibit sufficient entropy, improving the overall security of our scheme. The major contributions of the paper are as follows:
\begin{itemize}
    \item The concept of \textit{PUF Phenotype}, a \emph{new} classification approach, where PUF identity is considered to be the full externally observable PUF behaviour, including noise, without supplementary knowledge of PUF structure, physical properties, or environment. 
    
    \item DRAM Phenotype-based Authentication Network (DPAN), an ML-based group classification model, utilising model confidence to accurately identify and authenticate devices without the requirement for Helper Data Algorithms, Error Correcting Code or pre-filtering algorithms.  To the best of our knowledge, this work is the \emph{first} to consider a single ML model for multi-device PUF authentication. 
    
    \item Verification and analysis of the proposed scheme on \emph{our own} experimental dataset, using noisy DRAM-PUF responses collected under a broad range of environmental conditions. We provide this dataset as an open resource for the research community.
    
    \item Evaluation of classifier accuracy of DPAN for varying classifiers and group sizes up to 5 devices, demonstrating that good performance can be achieved without needing device specific classifiers. 
    
    \item Performance analysis of a DPAN implementation on a resource constrained system with analysis of power consumption, storage requirements, and execution time.
\end{itemize}

\subsection{Paper Organisation} Section \ref{preliminaries} provides a background on PUFs, DRAM-PUF and Machine Learning Modelling Attacks (ML-MA). In Section \ref{proposed_scheme} we introduce our proposed scheme, with our novel PUF Phenotype concept, dataset and classification methodology. Section \ref{results} provides the results and a discussion of our experiments using various classifiers and numbers of devices. We also provide a performance analysis of our model executed on a lightweight Raspberry Pi system in section \ref{experimental_analysis}. Finally, Section \ref{conclusion} concludes the work and identifies areas for further research.

\section{Preliminaries} \label{preliminaries}

\subsection{Physically Unclonable Functions}

A PUF is a hardware-rooted security primitive which provides an intrinsic identity, or ``fingerprint'', based on low level process variations in circuit components. As the individual process variation in each component is highly random, the distribution of variations across all components gives the circuit a unique identity. Those variations are a physical property and hard to duplicate. Therefore, that identity is strongly tied to the physical system. A PUF can fulfil a role similar to biometrics in human authentication systems, with the PUF identity being used to verify the identity of the parent device. PUFs can also be used as source of secret information, such as providing the secret key for a cryptographic algorithm, without that secret needing to be stored in secure memory. Unlike traditional cryptographic methods, PUFs aim to exploit unpredictable physical variations present in the hardware itself to extract entropy \cite{CCS:GCVD02}. 

A PUF may be considered as the function $r_n = PUF(c_n$) that accepts a set of $n$ input values (known as the Challenge) $C \in \{c_0,..,c_n\}$ from which are produced a corresponding output value/set of values (known as a the Response) $R \in \{r_0,..,r_n\}$. These Challenge-Response Pairs (CRPs) enable a PUF to act as a unique ``fingerprint'' of a device, which can be utilised by lightweight security schemes for secure key generation \cite{wenchao2014, devadas2007, gao2019, dram_retention_tehranipoor_et_al}, unique identification of devices \cite{gope_scalable, meng-day2016} and true random number generation \cite{BUCHOVECKA201733, vanderLeest2012}. Commonly, a verifier (most often a server) completes a secure enrollment phase wherein challenges and responses are monitored to generate a verifier-side database of CRPs for each PUF-equipped prover (most often a lightweight device). After distribution of the devices, during an authentication phase the server may issue a challenge from the CRP set to the device, to which the device generates a response on-the-fly using the supplied challenge as input to its PUF. As the correct PUF response can only be known, in theory, from the CRP database (known only to the server) or the PUF itself (accessible only by the device the PUF is a part of) returning the valid response proves knowledge of the PUF secret and hence the identity of the device.

\paragraph{PUF Types}
PUFs can be sorted into categories based on several criteria. A common division is Weak and Strong PUFs. This terminology can be somewhat misleading, as it does not refer to the security properties of the design but rather to the size of the CRP space (the full set of possible challenges and responses). A Weak PUF has a small CRP space (sometimes only a single CRP) which grows at a linear or polynomial rate with hardware size. A Strong PUF has a large CRP space which ideally grows exponentially with PUF size. Typically Weak PUFs are used for key generation, while Strong PUFs are used in CRP based authentication schemes. Another useful way of categorising PUFs is based on the evaluation method. In this case you have Extrinsic PUFs, where measuring the PUF response requires a method fully external to the PUF itself; Intrinsic PUFs, where the evaluation method is integrated into the PUF itself, often through the addition of circuitry designed for this purpose; and Software PUFs, where the evaluation method is integrated into the PUF and re-purposes existing circuitry, using software control mechanisms to perform the evaluation without design changes. Examples of Intrinsic PUFs include the Arbiter PUF (APUF) \cite{gassend2004}, XOR Arbiter PUF (XOR-APUF)  and Ring Oscillator PUF (RO-PUF) \cite{devadas2007}. These are custom hardware units which must be added to a device separately, increasing manufacturing costs and complexity. In addition, many PUFs of this type, especially so-called ``Strong'' designs, have been proven to be insecure to Machine Learning based Modelling Attacks (ML-MA) which aim to learn the underlying random behaviour and make accurate predictions of the PUF responses \cite{CCS:RSSDDS10, splittingInterpose, DelvauxJeroen2019MAoP}. A range of Software PUF designs have been proposed as an attractive solution to issues of cost and complexity in other PUF designs. Software PUFs utilise already existing hardware components on commodity devices, rather than requiring custom hardware installation. This lowers the barrier of entry for using a PUF in a given system. It also has the desirable property of making the PUF compatible existing IoT device designs and already manufactured devices via a firmware update. Memory-based PUFs such as Static Random Access Memory (SRAM), Dynamic Random Access Memory (DRAM) PUFs and more recently Processor-based PUFs \cite{maiti_schaumont_processor_puf, dscanpuf, belfast_processor_puf} are examples of such PUF designs that enable the generation of hardware-rooted identities using components that pre-exist on devices. Memory PUFs (SRAM, DRAM) in particular have very attractive characteristics for IoT security systems, due to their high storage capacity, ease of access and ubiquity in modern devices and appear to be (as of writing) secure against ML-MA.

\subsection{DRAM-PUF}
\begin{figure}[t!]
    \centering
    \makebox[0pt]{\includegraphics[width=50mm]{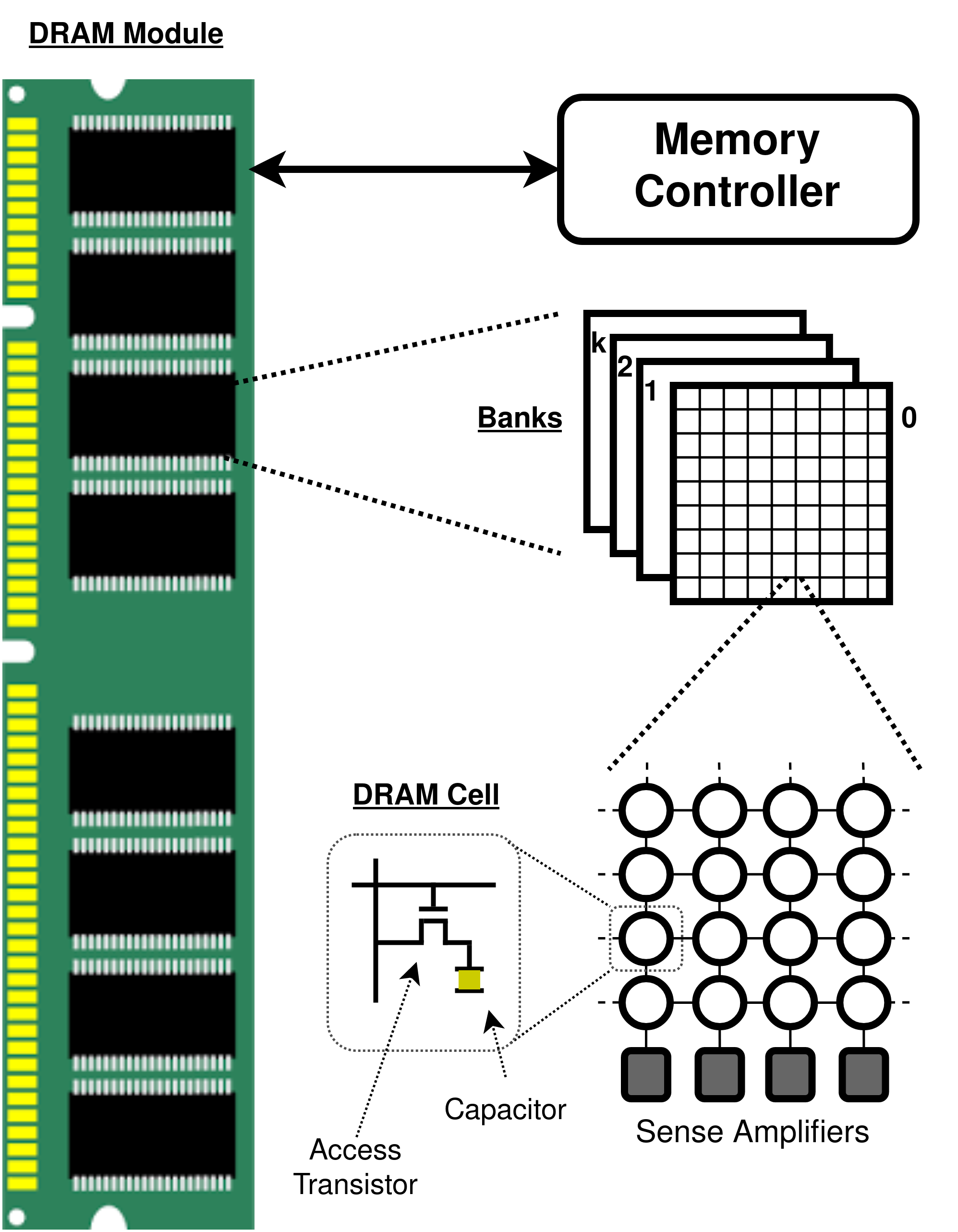}}
    \caption{DRAM Organisation \cite{fast_dram_puf}}
    \label{dram_diagram}
\end{figure}

Perhaps the most significant drawback of Software PUFs is that a given design can only be applied to systems which already contain a certain type of underlying hardware. If the hardware would need to be added just to facilitate a PUF, it defeats the purpose of using a Software PUF in the first place. Software Memory PUFs counter this problem by targeting an essential component, memory, which is found in almost all systems. The SRAM PUF has achieved great success and is one of the few PUF designs that sees widespread use in commodity devices \cite{xilinx_sram,intel_sram}. The more recent DRAM PUFs similarly target an even more ubiquitous memory type, DRAM, and thus are highly promising in terms of the range of potential target devices. At the simplest level DRAM consists of an array of transistor-gated capacitive cells (Figure \ref{dram_diagram}). The held value is represented in the charge value of the capacitor. Typically, a fully charged capacitor indicates binary 1, and a fully discharged capacitor binary 0. Most DRAM PUF designs make use of two properties of DRAM as an entropy measurement method. The first key property is that DRAM cells gradually lose their value over time due to charge leakage. This eventually leads to bit-flips in some cells and is why DRAM is considered a volatile memory type. To prevent data corruption, DRAM undergoes a periodic refresh cycle when active, reading and writing back cell contents to reset charge values. However, without the refresh cycle cells do not fail at a uniform rate. Within the cell arrays there is a highly random distribution of cells prone to rapid failure (i.e. are highly leaky) which is a product of process variation in the cell's physical structure. The first DRAM PUF proposals \cite{Liu2013, Keller2014, dram_retention_tehranipoor_et_al}, called DRAM Retention PUFs, use this as a means to generate the PUF response. The general process is as follows: first, a known pattern is written into the memory; next, the automatic refresh cycle is halted; the cells are allowed to begin failing, and after a fixed time period the refresh cycle is restored. This results in a new bit pattern in memory, the PUF response, based on the random distribution of fast failing cells across the memory block. This approach results in very high quality responses with strong security properties, but suffers from two major drawbacks: it cannot, generally, be performed during system runtime and it takes a long time, in the order of minutes, for enough cell failures to produce a good response.

The second key property of DRAM is that, being fundamentally a process of charging and discharging capacitors, there is an inherent latency in DRAM operations. To ensure correct behaviour there are built in delay timings for various aspects of internal operations. These timing parameters can be lowered from their factory defaults allowing for faster operation at the risk of causing instability. Too significant a reduction places the memory into an undefined behaviour state in which instructions may produce erroneous results. In PUF terms, the important factor is the process induced variation in the sources of internal latency, such as cell charge and discharge rate, line activation rate, the behaviour of sense amplifiers, etc. PUFs based on this property are called DRAM Latency PUFs \cite{dram_latency_kim_et_al,prelatpuf,fast_dram_puf}. By reducing or removing delays in the memory controller, internal DRAM operations are given insufficient time to execute resulting in timing errors. For example, severely reducing the \textit{tRCD} parameter (the required delay between opening a row and being able to access columns in it) and then attempting to read operations results in numerous read errors, with the resultant pattern being a product of both the values in the memory being read and process variations in the cell array and sense circuitry. This pattern of errors forms the PUF response, similar to the Retention PUF, and it is likewise the highly random distribution of process variations which make this a strong identifier for the memory. 
In DRAM Latency PUFs, unlike with DRAM Retention PUFs, the errors which comprise the PUF response are not actual cell failures. The errors occur from failure to accurately perform operations on the cell contents. If a read operation is used as the trigger, the cell contents remain entirely stable even as the read instructions return erroneous results. Due to this, PUFs of this type can in some circumstances be used in-runtime while maintaining system stability. They also allow response generation in much less time than Retention PUFs. A drawback of this approach is that responses tend towards high noise and are sensitive to environmental conditions, even comparable to other PUF designs. This calls for substantially more error correction and post-processing to produce a high quality response. DRAM Latency PUFs were chosen as a target for this work due to the degree of noise in individual responses and sensitivity to some environmental conditions they exhibit. They provide a good example of a PUF with strong potential for practical application but which can be limited in terms of performance by the energy and computational cost of error correction.

\subsection{PUF Vulnerabilities to ML}\label{puf_attacks}

Many strong PUFs have been shown over the years to be susceptible to machine learning modelling attacks (ML-MA). The first attacks which exploited the learnability of PUFs was performed by R{\"u}rhmair et al. in \cite{CCS:RSSDDS10} on Arbiter PUFs, XOR Arbiter PUFs, Lightweight Secure PUFs and Feed-Forward Arbiter PUFs. In their work, they collected a subset of available CRPs from each PUF type psuedorandomly following a standard normal distribution via an additive delay model. This set of CRPs formed the training dataset for various machine learning techniques in order to model new CRP data. More recently, helper data attacks utilising ML have been proposed to exploit the publicly known helper data used for error-correction to predict PUF responses. Streider et al.~\cite{ML_helper_data} demonstrate that public challenge data and helper data alone is sufficient to mount a successfully modelling attack on a PUF, even with as few as 800 helper data bits. Delvaux et al. ~\cite{DelvauxJeroen2014APPM} perform a helper data attack on Pattern Matching Key Generators, also exploiting their helper data. They demonstrate significant risk to full key recovery in an experiment on 4-XOR Arbiter PUFs.

\section{Proposed Scheme}\label{proposed_scheme}
In this section, we present our proposed scheme, which consists of three key parts: PUF Phenotype, the group-based authentication setting for our work and our DRAM Phenotype Authentication Network: DPAN.

\begin{figure*}[!ht]
    \centering
    \includegraphics[scale=0.08]{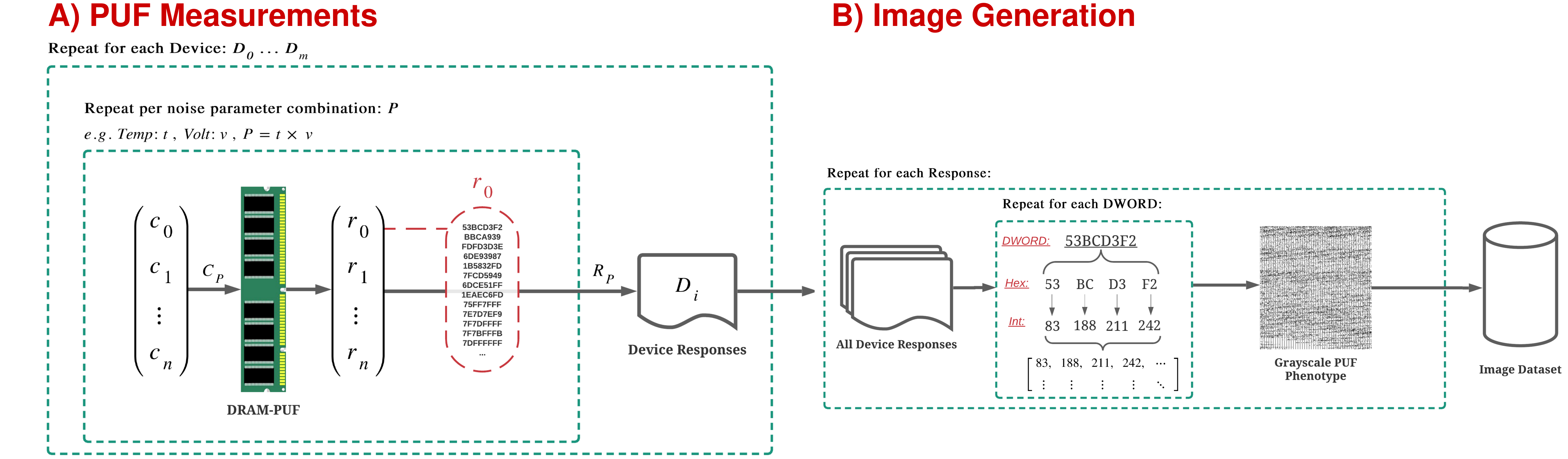}
    \caption{DRAM-PUF Phenotype Data Generation}
    \label{fig:DRAM_System}
\end{figure*}

\subsection{PUF Phenotype} \label{phenotype_section}

Biometric based identity is a field adjacent to and highly relevant to electronic PUFs. In a biological context there are two sets of connected identifying characteristics: Genotype and Phenotype. The genotype consists of a person's actual set of genetic instructions, while their phenotype is the set of externally observable traits which arise from the interaction between their genotype and the environment \cite{Taylor2017TheGD}. A specific example of genotype vs. phenotype would be as follows:\\

\begin{description}
\setlength\itemsep{2mm}
    \item \textbf{Genotype}: The unique sequence of DNA \textit{instructions} that exist to determine face shape.
    \item \textbf{Phenotype}: The resulting detectable expression (appearance) of the face, effected by environment.\\
\end{description}
    
While the phenotype is primarily derived from the genotype, the phenotype is \emph{not synonymous} with genotype, as the genotype cannot influence the effect environment has on the emergent properties of the phenotype. In biometric authentication, knowledge of the genotype is not required. If we consider the case of facial recognition, the computer vision system only has access to the \textit{externally observable characteristics} - the phenotype, plus environmental noise (incident light, angle, etc.). Despite this, such systems can perform authentication with a high degree of accuracy. Considering that a biometric is a form of Extrinsic PUF, this raises the question of whether, in the area of electronic PUFs, the costly on-device error correction and helper data (which may leak information about the PUF) which we often rely on is universally necessary for highly accurate authentication?

Drawing inspiration from the field of biometrics, we propose that an analogous distinction for electronic PUFs should be considered as an aspect of protocol design. That is, whether to consider the underlying structure of the PUF and try to remove noise so that only this underlying structure contributes to the identity, or whether to use a PUF Phenotype where both the PUF behaviour and the PUF response to environmental changes (i.e. noise) are treated as one set of identifying characteristics. The former is the default assumption in most PUF research, but there are advantages and disadvantages to both approaches. This is not to say that error correction based approaches should be discarded, rather that greater consideration should perhaps be given as to whether this default approach is optimal for a given solution. 

We express the idea of phenotype with regards to PUF in the generated images of our DRAM-PUF measurements. The physical organisation of the bit-cell matrix on the DRAM module is analogous to the genotype, while the PUF response image generated from that structure, including noise, is the expressed phenotype. The phenotype is not a visual representation of the PUF's physical structure (as was used in \cite{deeppuf, yueDRAM_CNN}, for example) but rather a structure-agnostic visualisation of the PUF identity. So long as the same method of image formation is used consistently, the phenotype can be generated from PUF data from any source, meaning knowledge of the PUF structure or even the type of PUF in use is irrelevant. Multiple PUF designs can be used in one system in this way, with a unified enrolment procedure, authentication database, and verifier. In addition, the phenotype includes environmental and inherent noise, reducing or negating the need for on-device error correction or the use of helper data. It is demonstrated in Section \ref{results} that using this approach devices can be authenticated with a high degree of confidence, even under highly variable environmental conditions and without any error correction mechanism. 

\subsubsection{Phenotype Dataset} \label{dataset}
We collected our own dataset of DRAM Latency PUF responses gathered from a test bed based on a Commodity Off-The-Shelf (COTS) AMD A series processor, controlling COTS DDR3 DRAM modules in DIMM form factor, using only software methods to acquire the PUF response\footnote{\textbf{The dataset used in this work is provided as an open resource for the use of the research community, and can be accessed at: \url{https://github.com/owenmillwood/Latency-DRAM-PUF-Dataset.git}}}.
Specifically, the Latency PUF method used was lowering the $t_{RCD}$ timing to 0 and attempting sequential reads of the target memory, using the method described in \cite{fast_dram_puf}. The testbed is effectively a small form factor desktop computer and runs a live version of Ubuntu Linux during experimentation. It should, therefore, closely reflect the performance and behaviour of the DRAM Latency PUF in a real-world scenario. Four QUMOX branded DDR3 DIMMs were tested, each comprising eight individual DRAM chips. For PUFs of this type, differences in the PUF response between manufacturers have previously been noted \cite{fast_dram_puf}, but when using appropriate post-processing or mixed input patterns the difference is relatively minor, meaning this dataset should be reasonably representative of DRAM Latency PUFs in general. The chips under test were characterised by targeting a representative physically contiguous section of 4Kb from the same starting point in each chip. The DRAM Latency PUF is data-dependent, so three input patterns were used in each experiment: 0xFF, 0x00, and 0x55. 

\paragraph{\textbf{DRAM-PUF Data Dependence}}\label{data_dependence}
Writing an 0xFF, 0x00 and 0x55 input pattern to the DRAM translates to an area of DRAM memory which (before lowering the $t_{RCD}$ timing parameter) contains all ones, zeroes and a mixed zero and one checkerboard pattern respectively. This provides three distinct challenge patterns that can be used per physical DRAM location for PUF operation. It is important to note that other randomly sequenced challenge patterns would not be suitable to generate PUF responses due to the various entropy sources of the DRAM-PUF and consequently the influence each source has on overall entropy. There are three sources to consider (in order of most to least influence on entropy):

\begin{itemize}
    \item The value stored in a cell, i.e. whether the cell is charging from the bit line (0 held in cell) or draining into it (1 held in cell). 
    
    \item The values in surrounding cells, due to charge leakage; Whether they are leaking charge out or drawing a small amount in influences the value of a given cell. The 0x55 checkerboard pattern is distinct from 0xFF and 0x00 because it alternates these influences, as oppose to all of one type.
    
    \item External influences such as line variation, sense amplifier variation and variation in memory controller, which is difficult to distinguish from environmental noise.
\end{itemize}

The influence on entropy of the surrounding cells is far smaller than the value in the cell itself for most cells, meaning randomly sequenced challenge patterns do not behave dissimilarly to the 0xFF, 0x00 and 0x55 patterns. Given an attacker obtaining knowledge of the 0xFF, 0x00 and Ox55 behaviour of the cells, it would be a trivial task to correctly predict the majority of the output to a given complex input pattern.

\begin{figure}[t]
    \centering
    \includegraphics[scale=0.100]{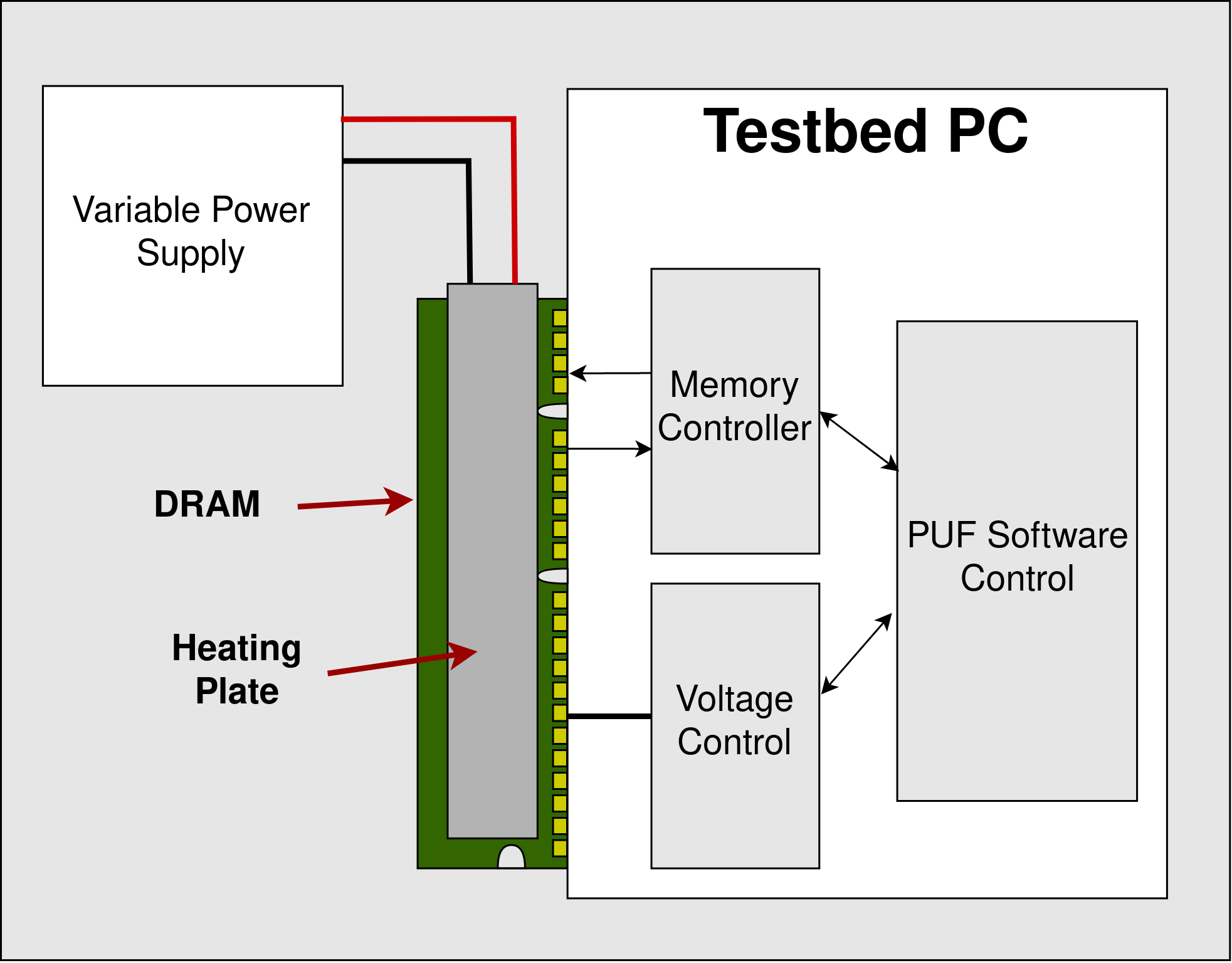}
    \caption{Experimental Setup for Dataset Generation}
    \label{fig:testbed}
\end{figure}

\paragraph*{\textbf{Noise}}
To measure the impact of environmentally-induced noise, the response for each input pattern was characterised at increasing temperatures, starting at a baseline of 20\degree C and increasing in 10\degree C intervals up to a maximum of 50\degree C (20\degree C, 30\degree C, 40\degree C, 50\degree C). This process was repeated at both a nominal DRAM voltage of 1.5v and a reduced 1.27v. The corner case of reduced voltage and high temperature produces the most noisy responses that can be expected from a PUF of this type and configuration in practice (discussed further in Section \ref{noise_section}). This provides a robust test for a noise tolerant authentication system such as the one presented in this work. The experimental setup for the collection of our dataset is shown in Figure \ref{fig:testbed}. The original measurements are formatted as a list of 32-bit DWORDs in hex format (eight characters e.g. FFFA3F6C) as read from the memory itself. This first process is described in part A of Figure \ref{fig:DRAM_System} and forms the scheme's Enrollment phase as described in Algorithm \ref{alg:enrollment}.\\

\subsubsection{Image Processing}
Each of these chip measurements were preprocessed into grayscale images to produce the dataset for training our model. Initially, we convert each pair of two hexadecimal digits (four pairs per DWORD line) into an integer ranging between 0-255. This integer denotes the intensity value (black to white) for a single pixel in the image, which is used to represent the response data visually. The size of each response measurement provides enough data to produce a grayscale image of size 220x200 pixels.
Each image was labelled with a class name corresponding to the device it was generated from. Rather than a numerical label, we chose the arbitrary labels: Alpha, Beta, Delta, Gamma and Epsilon for readability. These final DRAM-PUF response images are what we refer to as \textit{PUF Phenotype}. We name this process \textit{IMGEN}, which is described in part B of Figure \ref{fig:DRAM_System}.

\begin{figure}[tb]
\includegraphics[width=.35\linewidth]{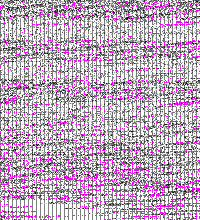}\hfill
\includegraphics[width=.35\linewidth]{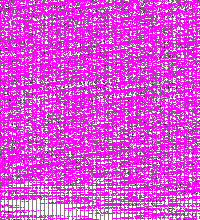}
\caption{Side-by-side Comparison of DRAM-PUF Measurements (noise highlighted in pink) \textbf{Left:} Response Measured in Ideal Environmental Conditions (5.95\% Noise) \textbf{Right:} Response Measured in Extreme Environmental Conditions (63.09\% Noise)}
\label{response_comparison}
\end{figure}

\subsubsection{Environmental Noise}\label{noise_section}
In any measurement from a physical process, some noise is expected. Even in ideal conditions most PUFs exhibit some noise, though usually to a degree which is usually manageable with error correction. Comparison of repeated measurements from the same device, chip, and challenge pattern reveal the effect environmental conditions have on noise in DRAM-PUF responses in our dataset. To demonstrate this, we compare an example of an ideal response with low noise against a response with high noise attributable by the measurement environment. We took a response measured in ideal conditions and compared it with another two repeated measurements with the same challenge (same device, chip and challenge pattern). One repeated measurement was taken in ideal conditions (20\degree C, normal voltage) and the other was measured in extreme conditions (50\degree C, low voltage)\footnote{\textbf{Both temperature and voltage effect noise in DRAM-PUF measurements. Measuring a response at a high temperature and lowered voltage represents the most extreme case of noise-inducing environment achievable by our dataset.}}. As these measurements were taken with the same challenge, if noise were not a factor the images would be identical. On comparison, the repeated measurement in ideal conditions showed a similarity of 94.05\%, demonstrating an intrinsic noise of 5.95\%. For the repeated measurement taken in extreme conditions, the noise effect is much more significant. On comparison, the similarity between the original measurement and measurement in extreme conditions was 63.09\%, producing 46.91\% noise in the response. This noise effect is shown visually in Figure \ref{response_comparison}.

%RECOMMENT FOR COLOUR
\begin{figure*}[!ht]
    \centering
    \makebox[0pt]{\includegraphics[width=142mm]{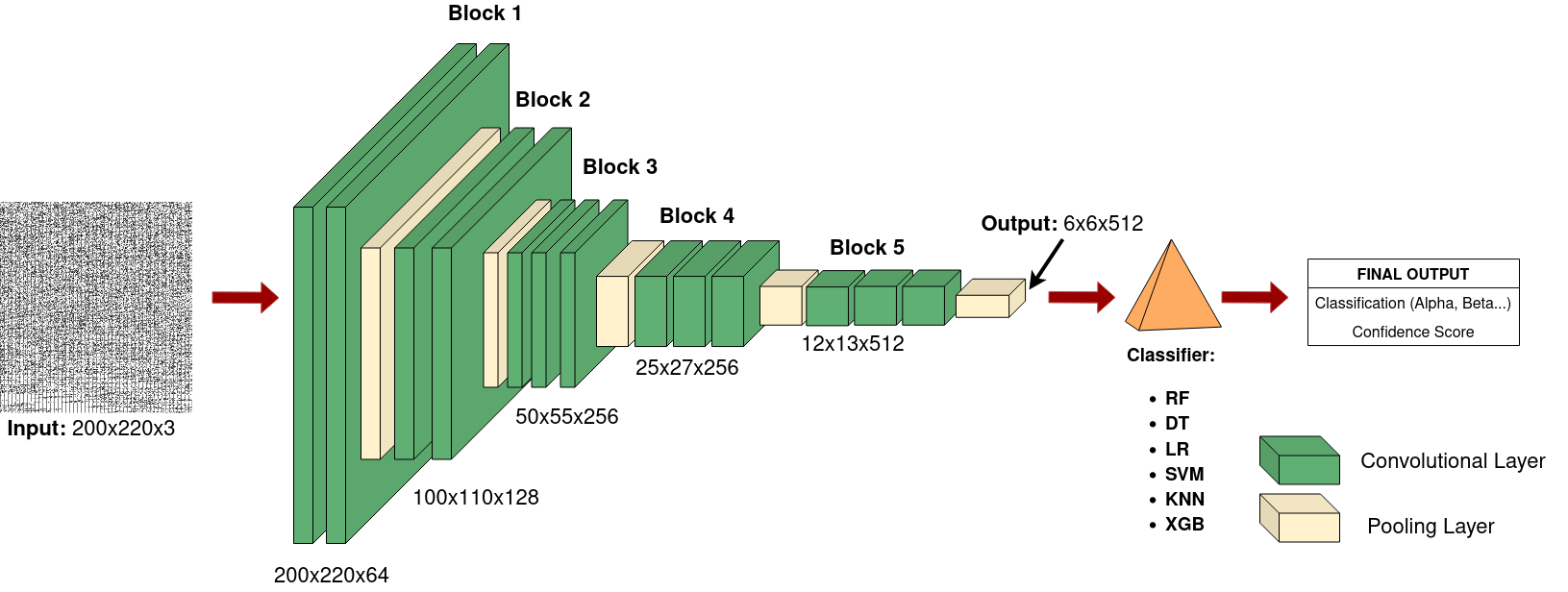}}
    \caption{DPAN Architecture}
    \label{vgg16diagram}
\end{figure*}

%RECOMMENT FOR GREYSCALE
% \begin{figure*}[!ht]
%     \centering
%     \makebox[0pt]{\includegraphics[width=180mm]{figures/DRAM_Model_Diagram_Grey.png}}
%     \caption{DPAN Architecture}
%     \label{vgg16diagram}
% \end{figure*}

%ENROLLMENT ALGORITHM
\begin{algorithm}[t]
\DontPrintSemicolon
  \centerline{\textbf{Dataset Generation}}
  \KwIn{$C \in \{c_0, c_1,...,c_n$\}: \textit{Challenge Data}}
  \KwOut{$X$: \textit{Phenotype Dataset}}
  \KwData{$P$: \textit{Environmental parameters} ($Temp$, $\mathit{Voltage}$)\\
  $m$: \textit{Number of devices}\\
  $n$: \textit{Number of challenges}
  }
\For{$i\gets0$ \KwTo $P$}
{
    \For{$j\gets0$ \KwTo $m$}
    {
        \For{$k\gets0$ \KwTo $n$}
        {
            Write Challenge Pattern from Start Location\;
            Set $t_{RCD}$ to 0\;
            $R_{ijk} \gets$ Perform Read operation on DRAM block\;
            $x$ $\gets$ IMGEN($R_{ijk}$)\\ \tcc{$x$: PUF Phenotype}
            Assign device label to $T$\;
            Store $x$ in $X$
        }
    }
}
   Return $X$\;
   \hrulefill
    
 \centerline{\textbf{Model Training}}
 \KwIn{$X$: \textit{PUF Phenotype Dataset}}
 \KwOut{$DPAN$: \textit{Trained model}\\
 $\tilde t$: \textit{Confidence threshold}}
 
 Split $X$ into train/test sets: $o$ \& $p$\;
 $F_a$ $\gets$ Fine-tune VGG16 using $o$\;
 \tcc{$F_a$: Training features}
 $F_b$ $\gets$ Fine tune VGG16 using $p$\;
 \tcc{$F_b$: Testing features}
 Train Classifier using $F_a$\;
 Test Classifier using $F_b$\;
 $DPAN \gets$ Combined trained VGG16 \& Classifier\;
 $\tilde t \gets$ Tune confidence threshold to zero false positives 
\Return $DPAN, \tilde t$
\caption{DPAN Enrollment Phases}
\label{alg:enrollment}
\end{algorithm}

\subsection{Intra-Group-Based Authentication Setting}

PUF-based authentication typically operates on a per-device basis. However, in some use cases it is desirable to not only confirm device identity, but to determine whether that device is a member of a group, for example, whether a given device is part of a group with elevated access privileges.
This can be done in existing PUF schemes layered atop of individual device authentication. This works well in scenarios where there is a highly resourced central verifier because it can be assumed the verifier can safely handle large amounts of data which must be kept secure from adversaries (e.g. CRP databases, group membership lists) and has substantial computational resources. However, when authenticating directly between IoT devices this poses significant challenges. In such a scenario storage space and computational resources are highly limited. Further, they operate in a less secure environment which limits what data can be safely stored on device, given that an insecure device could provide an entry point to a secure server for an attacker. We propose that using a computer vision based ML classification scheme operating on the PUF Phenotype data described above, it is possible to train a model to perform both individual PUF and group authentication. That is, when presented with a PUF response the model can determine first whether it is a group member and if so, which group member in particular, without any additional computational or storage requirements. Further, that this model is small and lightweight enough to allow direct device-to-device (intra-group) authentication in this manner. Using this, a single model may enable identification of multiple devices in a single model, saving the device from requiring an individual trained model to be stored per other device within its group.

To test this concept, we assume the following scenario:
\begin{itemize}
    \item There is a network of low resource IoT devices deployed in groups of up to 5 devices. The limitations of this method in regards to group size are discussed in Section \ref{results}
    \item Each device has a PUF which has been characterised prior to deployment as described in Section \ref{phenotype_section}.
    \item For each group a model has been trained prior to deployment using this data. This training is described in Section \ref{classification_scheme}.
    \item The model for each group is stored on each device in the network. The feasibility of this in terms of storage requirements is analysed in Section \ref{experimental_analysis}.
    \item Devices perform both individual and group authentication directly with no central verifier system. The computational and power costs of this are examined in Section \ref{experimental_analysis}.
\end{itemize}

\subsubsection{Confidence Decision Threshold}
When considering solutions for authentication, false positives and false negatives have different implications on the system. False negatives, where a legitimate entity is incorrectly denied authentication, simply impact system performance as a new handshake must be attempted. False positives however, where a malicious user is granted authentication, are far more important to consider. As an attacker would be free to query our system, there is nothing preventing them from sending a fabricated image to our model in an attempt to force it to classify it as one of the legitimate classes. In order to combat this type of attack, we utilise the model confidence to filter to authentication attempts. By applying a confidence threshold, received Phenotype images input to the model must reach a minimum level of confidence that must be achieved by the model to allow authentication. If a given classifier demonstrates a high confidence on average on correct classifications and low confidence on incorrect classifications, a threshold may be set to enable authentication. This is to say anecdotally, when a model outputs a low confidence, it is indicated that the model is unsure about whether the classification was correct. This provides two possible situations: one where the received image is from a legitimate device and one where the received image is fraudulent. In the first case, a low confidence score at worst causes a false negative (legitimate device is denied authentication). In the infrequent event that a legitimate device is rejected, they may simply initiate authentication again and succeed with a high probability. In the second case, an adversary would ideally not be able to produce their own responses which can allow for high model confidence on classification, enabling a higher true negative rate (fraudulent attempts correctly denied authentication). Taking this into account, a suitable \textit{confidence threshold} can be obtained from the average confidence score for both incorrect classifications of legitimate images and the arbitrary classification of fraudulent images. Such a threshold value would have to be obtained by PUF Phenotype images sent by the prover in order to be granted authentication. Increasing the confidence threshold has the effect of lowering false positives while potentially increasing false negatives. During enrollment, an ideal confidence threshold (minimum false negative rate) can be determined which eliminates false positives when testing on fraudulent data. This process is described in the Authentication Phase shown in Algorithm \ref{alg:authentication}. The ID check function which is applied in Algorithm \ref{alg:authentication} is shown in Algorithm \ref{alg:check}. We experimentally verify the confidence value property for each classifier with our results in Section \ref{confidence_section}.

%AUTHENTICATION ALGORITHM
\begin{algorithm}[!t]
\DontPrintSemicolon
 \centerline{\textbf{Authentication}}
 \KwIn{$x$: \textit{PUF Phenotype Image}}
 \KwData{$\tilde t$: \textit{Confidence threshold}}
 
 $\{UID|x\}\gets$ Current device receives authentication request\;
 \tcc{$UID$: Unique device identifier}
 \tcc{$x$: PUF Phenotype image }
 $Q \gets$ \textit{\textbf{Check}}$(UID)$\;
 \uIf{$Q == 0$}{
    Abort authentication \;
  }
  \Else{
    $\{\hat y, S\} \gets DPAN(x)$  \;
    \tcc{$\hat y$: Classification prediction\\ $S$: Confidence score}
    \uIf{\Not $\hat y == UID$}{
        Abort authentication\;
    }
    \ElseIf{$S < \tilde t$}
    {
        Abort authentication\;
    }
    \Else{
       Phenotype authenticity passed\;
    }
  }
\caption{DPAN Authentication Phase}
\label{alg:authentication}
\end{algorithm}

%CHECK ALGORITHM
\begin{algorithm}[t]
 \centerline{\textbf{Check}}
 \KwIn{$ID$: \textit{Recieved Device Identifier}}
 \KwOut{$\delta \in \{0,1\}$: \textit{ID found decision}}
 \KwData{$DB_{UID}$: \textit{UID List}\\
 $q$: Size($DB_{UID}$)}
 
 \ForEach{$DB_{UID} \in \{UID_0,...,UID_q\}$}
    {
        \uIf{$UID == ID$}
            {
                $\delta \gets 1$\;
                \Return $\smallskip \delta$\; 
            }
        \ElseIf{\Not EOF}
        {
         \tcc{EOF: End Of File}
            \Continue
        }
        \Else
        {
            $\delta \gets 0$\;
            \Return $\smallskip \delta$\;
        }
    }
\caption{UID Check Function}
\label{alg:check}
\end{algorithm}  

\subsection{DRAM Phenotype-based Authentication Network (DPAN)} \label{classification_scheme}

We use a Convolutional Neural Networks (CNN) in a popular model architecture consisting of a feature extractor, whose output is fed into a classifier as input. The feature extractor unit is a VGG16 CNN, which uses pooling layers to output lower-dimensional representations of the PUF response Phenotype images. This stage is effective as it enables the classifier to be trained on more appropriate features of the PUF Phenotype image data.

\subsubsection{Feature Extraction - Modified VGG16}
We utilised a modified VGG16 CNN framework as our feature extractor unit due to its combined simplicity, efficiency and powerful capability as an image feature extractor in comparison to other well-known CNNs (AlexNet, ResNet) \cite{vgg16_simonyan}. The architecture of the VGG16 used is shown in Figure \ref{vgg16diagram}. The network consists of 5 blocks of convolutional layers, which aggregate local information, each followed by a pooling layer performing dimensionality reduction, with a total of 13 convolutional layers and 5 pooling layers. All the convolutional layers have 3×3 filters, and the pooling size is 2×2. In a standard (unmodified) VGG16 model, there are three dense (fully connected) layers, which perform the final classification of the features extracted from the initial convolutional and pooling layers. As the name suggests, fully connected layers consist of nodes, each with its own connection to every node in the following layer. These layers enable extremely powerful function approximation, provided sufficient training data is available to fine-tune the layers. Naturally, these layers are the most data-intensive and account for the majority of storage requirement for a neural network, leaving the network at around 0.5GB. In our proposed scheme, however, we replace these dense layers with lighter standard ML classifiers. As a result, we replace the three fully connected layers with a 1×1 average pooling layer to convert the original VGG16 from classification to only feature extraction. The original 200x220x3 input image is therefore output by the feature extractor as a 6x6x512 feature vector. Consequently, the lightweight VGG16 utilised by our scheme is only 57MB, reducing the storage and memory requirement by a factor of almost 10. This broadens the available applications of DPAN, enabling storage on many restricted devices, which is further discussed in Section \ref{experimental_analysis}. To compare the performance of the modified VGG16 with a standard VGG16, we provide benchmark classification results from an unmodified full VGG16 model with the dense layers intact as a feature extractor, which is discussed further in Section \ref{results}.\\

\subsubsection{Tested Classifiers}
%SVM, LR, SGD, RF, KNN, XGBoost
The features output from the VGG16 model is utilised to train a standard ML classifier. To determine the best performing classifier for our application, we tested multiple popular classifiers on our data, each having passed through the same modified VGG16 CNN. As our data is labelled, we chose six supervised learning techniques to experiment with: XGBoost, Support Vector Machine, K-Nearest Neighbours, Random Forest, Logistic Regression and Decision Tree. This distribution of techniques represents a broad spectrum of standard classification techniques, including faster, more simple approaches such as K-nearest neighbours, alongside state-of-the-art ensemble methods such as XGBoost. For the interested reader, a brief description of each classifier has been included in Appendix C of the supplementary material.\\

\subsubsection{Training and Evaluation}\label{training_and_eval}
We tested the modified VGG16 in combination with each classifier on the test data and reported \textit{Accuracy} and \textit{F1-score} as our performance metrics.
For ablation study, we also compare our approach based on the lightweight modified VGG16 against a full VGG16 model trained with the dense layers present to perform classification for groups of 3, 4 and 5 devices respectively. Finally, during training we performed $K$-fold ($K=5$) cross validation which ensured we could test our models across the entire distribution of the dataset as oppose to a standard train/test split, where the model only ever sees a specific portion of the dataset\footnote{\textbf{For the interested reader, a more detailed explanation for each evaluation metric is provided in Appendix D of the supplementary material.}}.\\

\subsubsection{Hyperparameter Tuning}

Before finding our final model results, we performed hyperparameter optimisation for each classifier to enhance the performance of our models for our specific classification task. We performed a randomised grid search through a predetermined list of each available hyperparameter per classifier. During this phase, we used $K$-fold ($K=5$) cross validation as mentioned in Section \ref{training_and_eval}. when training each model with each hyperparameter combination. The specific hyperparameters chosen for each classifier can be found at Appendix E in the supplementary material.

\section{Results and Discussion}\label{results}

In this section we evaluate the performance of each tested classifier in its classification performance using both the full VGG16 feature extractor and the modified lightweight VGG16 feature extractor. We also evaluate the confidence scores output for each model and discuss the ability of each classifier to identify legitimate phenotype inputs from fraudulent ones. Each classifier was tested for each three to five devices as discussed in Section \ref{training_and_eval}. Tables \ref{tab:full_vgg_classification_results} and \ref{tab:lightweight_classification_results} display the accuracy, F1 score, cross-validated mean accuracy and cross-validated standard deviation for each tested classifier.

%FULL VGG16 WITH CLASSIFIERS
\begin{table*}[ht]
\scriptsize
\begin{threeparttable}
\setlength\tabcolsep{0pt}
\centering
\caption{Classification Results with Full VGG16 Feature extractor}
\label{tab:full_vgg_classification_results}
\begin{tabular*}{\textwidth}{@{\extracolsep{\fill}}p{1.2cm} cccccc@{\hskip 0.03in} cccccc@{\hskip 0.03in} cccccc@{\hskip 0.03in} cccccc}
\toprule
\multirow{2}{*}{\parbox{1cm}{\centering\textbf{No. of Devices}}}  &

 \multicolumn{6}{c}{\textbf{Accuracy}} &\multicolumn{6}{c}{\textbf{F1 Score}}
     &\multicolumn{6}{c}{\textbf{CV\tnote{**} Mean Accuracy}}
     &\multicolumn{6}{c}{\textbf{CV Standard Deviation}}
     \\
\cmidrule{2-7} \cmidrule{8-13} \cmidrule{14-19} \cmidrule{20-25}

    & \textbf{SVM}\tnote{*} & \textbf{LR} & \textbf{DT} & \textbf{KNN} & \textbf{RF} & \textbf{XGB}
    & \textbf{SVM} & \textbf{LR} & \textbf{DT} & \textbf{KNN} & \textbf{RF} & \textbf{XGB}
     & \textbf{SVM} & \textbf{LR} & \textbf{DT} & \textbf{KNN} & \textbf{RF} & \textbf{XGB}
      & \textbf{SVM} & \textbf{LR} & \textbf{DT} & \textbf{KNN} & \textbf{RF} & \textbf{XGB} \\
\midrule
\textbf{3x Devices} & \underline{\textbf{0.957}} & \underline{\textbf{0.957}} & 0.793 & 0.922 & 0.948 & \underline{\textbf{0.957}} &
\underline{\textbf{0.957}} & \underline{\textbf{0.957}} & 0.796 & 0.923 & 0.949 & \underline{\textbf{0.957}} &
0.993 & \underline{\textbf{0.996}} & 0.898 & \underline{\textbf{0.996}} & \underline{\textbf{0.996}} & \underline{\textbf{0.996}} &
0.009 & \underline{\textbf{0.005}} & 0.019 & \underline{\textbf{0.005}} & \underline{\textbf{0.005}} & \underline{\textbf{0.005}} 

\\
\textbf{4x Devices} & 0.955 & \underline{\textbf{0.968}} & 0.844 & 0.942 & 0.955 & 0.948 &
0.955 & \underline{\textbf{0.968}} & 0.846 & 0.943 & 0.955 & 0.948 &
0.971 & \underline{\textbf{0.985}} & 0.871 & 0.980 & 0.984 & 0.984 &
\underline{\textbf{0.025}} & 0.030 & 0.038 & 0.028 & 0.029 & 0.029 
\\
\textbf{5x Devices} & \underline{\textbf{0.938}} & 0.917 & 0.651 & 0.922 & 0.891 & 0.854 &
\underline{\textbf{0.938}} & 0.917 & 0.651 & 0.922 & 0.890 & 0.854 &
\underline{\textbf{0.987}} & 0.982 & 0.853 & 0.976 & 0.978 & 0.978 &
\underline{\textbf{0.023}} & 0.033 & 0.065 & 0.031 & 0.035 & 0.035 
\\
\bottomrule
\end{tabular*}
\smallskip
\scriptsize
\begin{tablenotes}
\RaggedRight
\item[*]    \textbf{DT}: Decision Tree; \textbf{KNN}: k-Nearest Neighbors;
\textbf{LR}: Logistic Regression;
\textbf{XGB}: XGBoost;
\textbf{RF}: Random Forrest;
 and
\textbf{SVM}: Support Vector Machine; ** \textbf{CV:} Cross-Validation
\end{tablenotes}
\end{threeparttable}
\end{table*}

%LIGHTWEIGHT VGG WITH CLASSIFIERS
\begin{table*}[ht]
\scriptsize
\begin{threeparttable}
\setlength\tabcolsep{0pt}
\centering
\caption{Classification Results with Lightweight VGG16 Feature extractor}
\label{tab:lightweight_classification_results}
\begin{tabular*}{\textwidth}{@{\extracolsep{\fill}}p{1.2cm} cccccc@{\hskip 0.03in} cccccc@{\hskip 0.03in} cccccc@{\hskip 0.03in} cccccc}
\toprule
\multirow{2}{*}{\parbox{1cm}{\centering\textbf{No. of Devices}}}  &

 \multicolumn{6}{c}{\textbf{Accuracy}} &\multicolumn{6}{c}{\textbf{F1 Score}}
     &\multicolumn{6}{c}{\textbf{CV Mean Accuracy}}
     &\multicolumn{6}{c}{\textbf{CV Standard Deviation}}
     \\
\cmidrule{2-7} \cmidrule{8-13} \cmidrule{14-19} \cmidrule{20-25}

    & \textbf{SVM} & \textbf{LR} & \textbf{DT} & \textbf{KNN} & \textbf{RF} & \textbf{XGB}
    & \textbf{SVM} & \textbf{LR} & \textbf{DT} & \textbf{KNN} & \textbf{RF} & \textbf{XGB}
     & \textbf{SVM} & \textbf{LR} & \textbf{DT} & \textbf{KNN} & \textbf{RF} & \textbf{XGB}
      & \textbf{SVM} & \textbf{LR} & \textbf{DT} & \textbf{KNN} & \textbf{RF} & \textbf{XGB} \\
\midrule
\textbf{3x Devices} & 0.983 & 0.983 & 0.957 & 0.983 & 0.983 & \underline{\textbf{0.991}} &
0.983 & 0.983 & 0.957 & 0.983 & 0.983 & \underline{\textbf{0.991}} &
0.983 & \underline{\textbf{0.998}} & 0.989 & \underline{\textbf{0.998}} & 0.996 & 0.996 &
0.035 & \underline{\textbf{0.004}} & 0.022 & \underline{\textbf{0.004}} & 0.009 & 0.009 

\\
\textbf{4x Devices} & 0.974 & \underline{\textbf{0.981}} & 0.968 & 0.968 & \underline{\textbf{0.981}} & 0.961 &
0.975 & \underline{\textbf{0.981}} & 0.968 & 0.969 & \underline{\textbf{0.981}} & 0.962 &
0.975 & \underline{\textbf{0.995}} & 0.974 & \underline{\textbf{0.995}} & \underline{\textbf{0.995}} & \underline{\textbf{0.995}} &
0.049 & \underline{\textbf{0.010}} & 0.048 & \underline{\textbf{0.010}} & \underline{\textbf{0.010}} & \underline{\textbf{0.010}}  
\\
\textbf{5x Devices} & 0.979 & 0.979 & 0.927 & 0.974 & \underline{\textbf{0.984}} & 0.974 &
0.979 & 0.979 & 0.927 & 0.974 & \underline{\textbf{0.984}} & 0.974 &
0.959 & 0.992 & 0.973 & \underline{\textbf{0.993}} & 0.991 & 0.991 &
0.075 & \underline{\textbf{0.010}} & 0.048 & \underline{\textbf{0.010}} & 0.015 & 0.015 
\\
\bottomrule
\end{tabular*}
\smallskip
\scriptsize
\end{threeparttable}
\end{table*}

%CONFIDENCE SCORES
\begin{table*}[ht]
\scriptsize
\begin{threeparttable}
\setlength\tabcolsep{0pt}
\centering
\caption{Confidence Scores}
\label{tab:confidence_results}
\begin{tabular*}{\textwidth}{@{\extracolsep{\fill}}p{1.2cm} cccccc@{\hskip 0.03in} cccccc@{\hskip 0.03in}  cccccc@{\hskip 0.03in} cccccc}
\toprule
\multirow{2}{*}{\parbox{1cm}{\centering\textbf{No. of Devices}}}  &
 \multicolumn{6}{c}{\textbf{Mean Correct Confidence}} &\multicolumn{6}{c}{\textbf{Mean Incorrect Confidence}}
 &\multicolumn{6}{c}{\textbf{Max Adversary Confidence}}
 &\multicolumn{6}{c}{\textbf{FN--FP (\%)}\tnote{\textdagger\textdagger}}
     \\
\cmidrule{2-7} \cmidrule{8-13} \cmidrule{14-19} \cmidrule{20-25}
    & \textbf{SVM} & \textbf{LR} & \textbf{DT}\tnote{\textdaggerdbl\textdaggerdbl} & \textbf{KNN} & \textbf{RF} & \textbf{XGB}
    & \textbf{SVM} & \textbf{LR} & \textbf{DT} & \textbf{KNN} & \textbf{RF} & \textbf{XGB} 
    & \textbf{SVM} & \textbf{LR} & \textbf{DT} & \textbf{KNN} & \textbf{RF} & \textbf{XGB}
      & \textbf{SVM} & \textbf{LR} & \textbf{DT} & \textbf{KNN} & \textbf{RF} & \textbf{XGB} \\
\midrule
\textbf{3x Devices} & 0.978 & \underline{\textbf{0.992}} & - & 0.981 & 0.975 & 0.761 &
0.800 & 0.765 & - & 0.778 & 0.746 & \underline{\textbf{0.707}} &
0.748 & 0.984 & - & 0.667 & 0.793 & \underline{\textbf{0.395}} &
\underline{\textbf{2.6--0}} & 10.3--70 & - & 7.8--0 & 9.5--0 & 3.5--0
\\
\textbf{4x Devices} & 0.952 & \underline{\textbf{0.979}} & - & 0.985 & 0.959 & 0.713 &
\underline{\textbf{0.546}} & 0.834 & - & 0.778 & 0.783 & 0.596 &
\underline{\textbf{0.355}} & 0.945 & - & 1.000 & 0.548 & 0.652 &
\underline{\textbf{3.3--0}} & 9.1--0 & - & 7.1--20 & 5.8--0 & 10.4--0 
\\
\textbf{5x Devices} & 0.927 & \underline{\textbf{0.979}} & - & 0.974 & 0.936 & 0.674 &
0.535 & 0.800 & - & 0.667 & 0.621 & \underline{\textbf{0.450}} &
\underline{\textbf{0.390}} & 0.900 & - & 0.778 & 0.664 & 0.639 &
\underline{\textbf{2.1--0}} & 8.9--0 & - & 8.3--0 & 9.4--0 & 17.2--0
\\
\bottomrule
\end{tabular*}
\smallskip
\scriptsize
\begin{tablenotes}
\RaggedRight
\item[\textdagger\textdagger] \textbf{FN:} False Negative; \textbf{FP:} False Positive; \textdaggerdbl\textdaggerdbl  Decision Tree not applicable due to inability to determine confidence
\end{tablenotes}
\end{threeparttable}
\end{table*}

%EXPERIMENTAL ANALYSIS
\begin{table*}[ht]
\scriptsize
\begin{threeparttable}
\setlength\tabcolsep{0pt}
\centering
\caption{DPAN Device Overhead}
\label{tab:device_overheads}
\begin{tabular*}{\textwidth}{@{\extracolsep{\fill}}p{1.2cm} cccccc@{\hskip 0.03in} cccccc@{\hskip 0.03in} cccccc}
\toprule
\multirow{2}{*}{\parbox{1cm}{\centering\textbf{No. of Devices}}}  &
 \multicolumn{6}{c}{\textbf{Max Power Overhead (\textit{VA})}} &\multicolumn{6}{c}{\textbf{Storage Requirement (\textit{KB})}} &\multicolumn{6}{c}{\textbf{Execution Time (\textit{Seconds})}}
     \\
\cmidrule{2-7} \cmidrule{8-13} \cmidrule{14-19}
    & \textbf{SVM} & \textbf{LR} & \textbf{DT} & \textbf{KNN} & \textbf{RF} & \textbf{XGB}
    & \textbf{SVM} & \textbf{LR} & \textbf{DT} & \textbf{KNN} & \textbf{RF} & \textbf{XGB}
     & \textbf{SVM} & \textbf{LR} & \textbf{DT} & \textbf{KNN} & \textbf{RF} & \textbf{XGB} \\
\midrule
\textbf{3x Devices} & 5.97 & 5.84 & 5.79 & 6.04 & \underline{\textbf{4.85}} & 5.76 &
359 & 13 & \underline{\textbf{2}} & 925 & 276 & 108 &
5.76 & \underline{\textbf{5.72}} & 6.61 & 5.80 & 6.79 & 5.67 
\\
\textbf{4x Devices} & \underline{\textbf{4.78}} & 6.00 & 4.94 & 4.86 & 4.94 & 5.94 &
948 & 17 & \underline{\textbf{3}} & 1,234 & 648 & 170 &
6.98 & 6.18 & \underline{\textbf{5.96}} & 6.33 & 7.21 & 6.68
\\
\textbf{5x Devices} & 5.86 & 5.80 & 5.90 & \underline{\textbf{5.72}} & 5.78 & 6.20 &
1,237 & 21 & \underline{\textbf{3}} & 1,541 & 1,241 & 261 &
6.51 & \underline{\textbf{6.16}} & 6.28 & 6.42 & 7.43 & 6.26  
\\
\bottomrule
\end{tabular*}
\smallskip
\scriptsize
\end{threeparttable}
\end{table*}

\subsection{Full VGG16 Feature Extractor Benchmark}\label{full_classification_results}

The results in Table \ref{tab:full_vgg_classification_results} demonstrate varying levels of success in classification of each device between each classifier. The LR classifier displayed the best performance across most of the tests, with a classification accuracy of 95.7\%, 96.8\% and 91.7\% for three, four and five devices, respectively. SVM also preformed very well in each test, showing slightly better performance on five devices with an accuracy of 93.8\%. While the RF and XGB classifiers performed comparably for fewer devices, each of these classifiers suffered more significantly for five devices with accuracies down to 89.1\% and 85.4\% respectively. The DT classifier demonstrated a vastly worse performance than each other classifier, showing the worst performance across each number of devices with classification accuracy as low as 65.1\%. The highest achieved result for each number of devices was 95.7\%, 96.8\% and 93.8\%. Intuitively, as the number of classes (devices) increases, the performance drops slightly on average due to the increased complexity of the models.

\subsection{Proposed Lightweight VGG16 Feature Extractor}

Applying the lightweight VGG16 feature extractor showed a far greater performance across all classifiers in comparison to the full VGG16 feature extractor. The results shown in Table \ref{tab:lightweight_classification_results} indicate a strong ability for most models to accurately classify the origin of each PUF Phenotype in the testing set. Here again, the LR and SVM models performed the most consistently across the tests for each number of devices, correctly classifying almost all Phenotype images with between 97.9\% to 98.3\% accuracy. For the three device model, XGB performed exceptionally well, correctly classifying 99.1\% of Phenotypes correctly. The XGB model did, however, suffer poorer results slightly with the increased number of devices. The RF classifier performed almost identically to the LR classifier; however, it performed the best for the higher number of devices, achieving an accuracy of 98.4\% for five devices. Overall, the LR and RF models were the highest overall performing classifiers. The confusion matrices for these two classifiers for each number of devices can be seen in Figure \ref{confusion matrices}\footnote{\textbf{The confusion matrices for all experiments are provided in Appendices A and B of the supplementary material.}}.

\begin{figure}[t!]
\centering
\includegraphics[width=81mm]{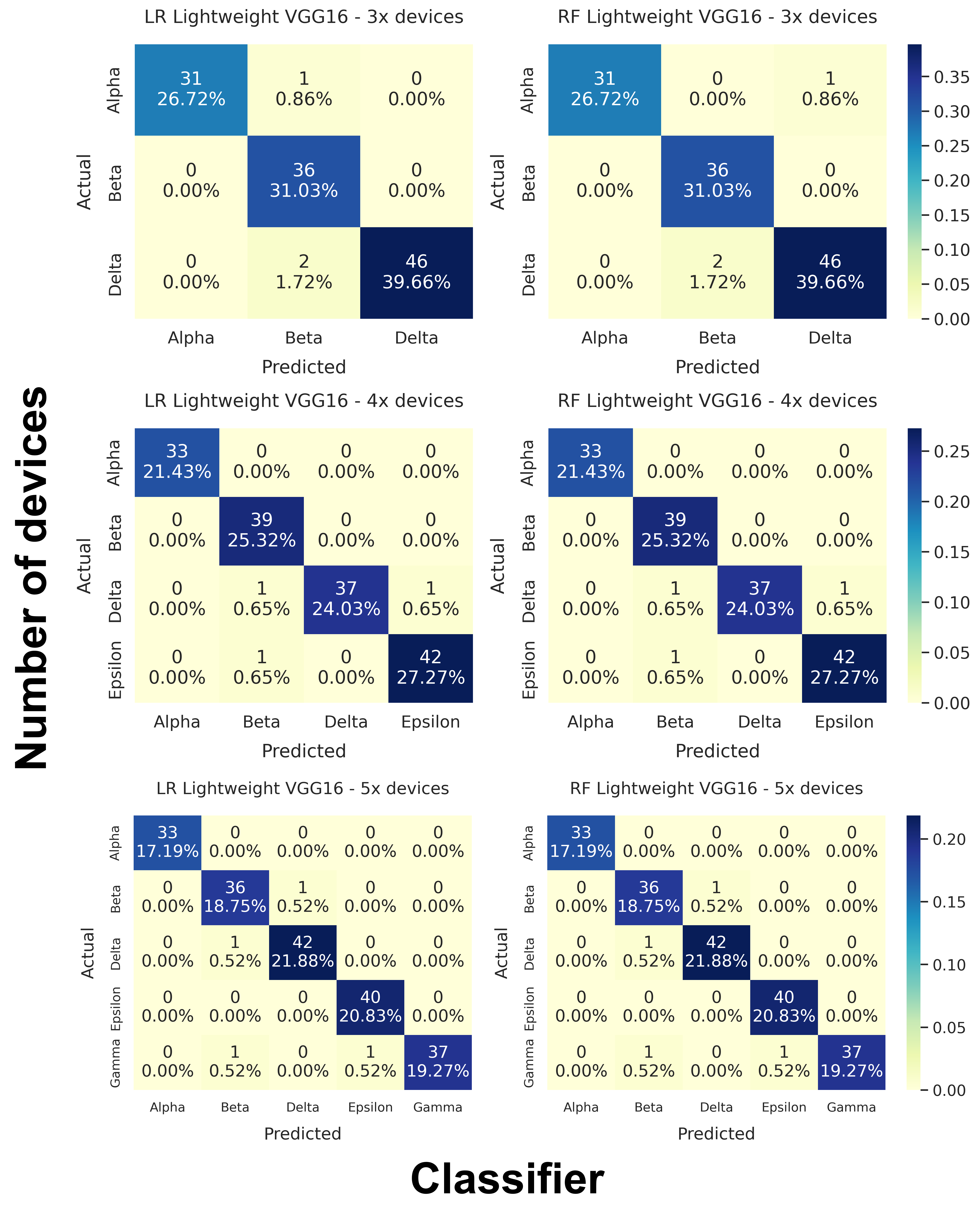}
\caption{Confusion Matrices of the Best Performing Classifiers for 3, 4 and 5 Devices}
\label{confusion matrices}
\end{figure}

\subsection{Model Confidence}\label{confidence_section}

Table \ref{tab:confidence_results} shows the average confidence values for each correctly and incorrectly classified legitimate images, the maximum confidence over a set of ten fraudulent randomly generated images and finally the False Negative and False Positive rates determined by a tuned confidence value for each classifier\footnote{\textbf{The tuned confidence thresholds for each model have been included in Appendices F of the supplementary material.}}. We see that both the SVM and RF classifiers perform strongly when identifying fraudulent images. SVM performed exceptionally well with 4 to 5 devices, with maximum confidence outputs of 0.355 and 0.390 respectively. When using a tuned confidence score to authenticate each test sample, SVM showed the best performance with no false positives for fraudulent samples and a maximum of 3.3\% false negative rate across legitimate test samples. RF also performed strongly, however suffered from a slightly higher false negative rate of between 5.8\% and 9.4\%. While LR proved an excellent classifier of legitimate images, it performed poorly in identifying fraudulent images with a minimum confidence score of 0.900 (5 devices). XGB performed strongly in removing false positive results, however displayed a high false negative rate, which translates to poorer overall system performance.

\subsection{Device Overhead}\label{experimental_analysis}

\begin{figure}[!t]
    \centering
    \makebox[0pt]{\includegraphics[width=70mm]{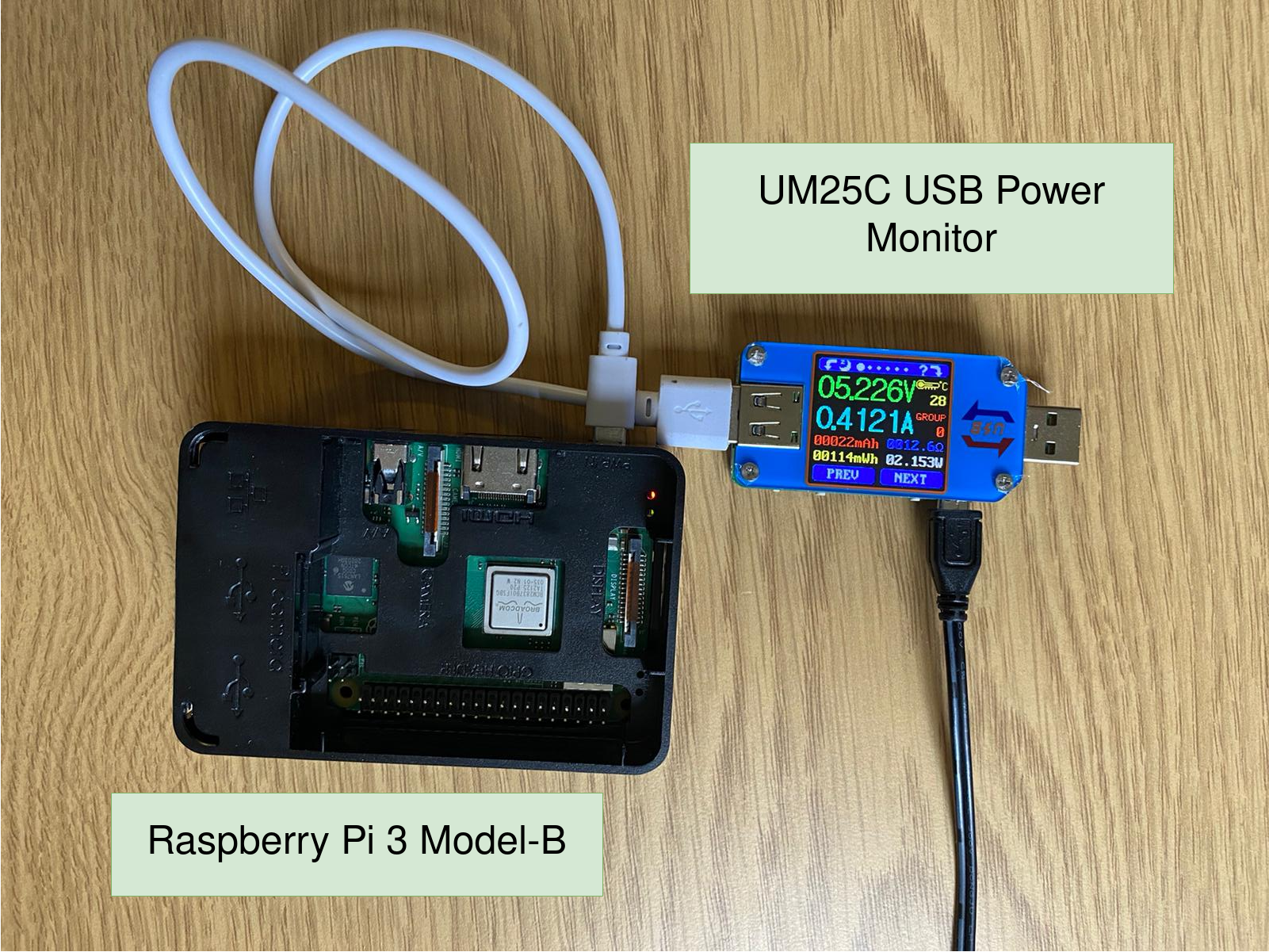}}
    \caption{Raspberry Pi Setup for Device Overhead Analysis}
    \label{rpi_figure}
\end{figure}

We finally performed an analysis of model execution for each classifier on a Raspberry Pi 3 Model-B to simulate performance in a resource constrained system (Figure \ref{rpi_figure}). Table \ref{tab:device_overheads} provides the power overhead and execution time for a single image classification for each model alongside the storage requirements for each. The size of the modified VGG16 feature extractors were 57,589\textit{KB}, 57,591\textit{KB} and 57,593\textit{KB} for the three, four and five device models respectively. The feature extractor size has been omitted from the table for ease of reading as each classifier uses the same lightweight feature extraction model. To determine the maximum power overhead, we monitored the power consumption of the Raspberry Pi in an idle state for three hundred seconds to establish a baseline power consumption of the device. During idle, the Raspberry Pi consumes on average $3.371$ Volt Amperes, therefore we considered any value over this during model execution to be the power requirement overhead of for each model. In terms of maximum power consumption, no single classifier performed the best over each of the three, four and five device models, with RF demonstrating the lowest overhead for the three device classification, SVM for four devices and KNN for five devices. Execution time was more consistent, with the LR model performing the best in terms of execution times compared to the other models. RF overall had a higher required execution time for classification than the other classifiers, however this was not significant when considering its effectiveness in terms of accuracy and Phenotype authentication as discussed previously.

\section{DPAN Security Considerations}

We address a set of possible security threats/concerns for the proposed system. We assume an adversary has physical access to the DRAM-PUF device, and are able to access and inspect the trained classification model stored on device. As a formal protocol is out of scope of this work, we assume protocol level security to be provided supplementary to the classification system to regard attacks such as replay, message tampering, denial of service, desynchronisation etc.
\paragraph*{\textbf{Inference of PUF Behaviour}}
An adversary may attempt to gain learn the PUF response data/entropy at some stage of the authentication phase. Usually, it would be necessary to consider an adversary looking to collect some helper data used for error correction in order to learn about the PUF CRP behaviour; in the proposed scheme, however, no helper data is required at any point to actively correct errors in the responses themselves, negating this security threat entirely. This leaves one threat to consider, being that of an adversary attempting to learn some information about the DRAM-PUF behaviour from the classification model itself. Due to the black-box nature of trained neural networks, and the fact that the input and output are already publicly known, an inspection of the trained VGG16 or classifier would reveal no useful information to allow an adversary to infer the PUF behaviour.
\paragraph*{\textbf{Response Guessing}}
An adversary may attempt to brute force query the classification model to try to discover which images produce an acceptable confidence score to allow authentication. This would require a random guess of a response close enough to the hamming distance threshold that also produces an acceptable confidence value when input to the model.   This attack is not possible as an attacker must guess the correct integer from 0-255 for each one of 44,000 pixels in the image data, in addition to the guessed pixel value being in an order correct enough for the model to extract comparable features to the legitimate images. We also experimentally verify this security property in Section \ref{confidence_section}, as when utilising an appropriate confidence threshold, DPAN is able to distinguish fraudulent responses to mitigate false authentication from adversarial images entirely. Finally, an attacker who has knowledge about the algorithms and class labels used in DPAN may attempt to collect some device responses to train their own new classification models and attempt to learn some information to compromise the PUF and/or authentication system. This attack is also not economical for an attacker as training a new model is also a black-box process and does not enable an adversary to learn particular new inputs that could be used to deceive a legitimate model. An extremely similar if not identical classification model would need to be trained (which is difficult to confirm by the attacker) and brute force queried with fraudulent images to identify patterns that could potentially fool the real DPAN model.
\paragraph*{\textbf{Poisoning Attack}}
An adversary may attempt to send multiple noisy data to a device to impact the future performance of the model. This type of attack is not possible as poisoning attacks only effect online models which are continuously trained after deployment to remain up to date with continuous data and avoid concept drift. As training only occurs during secure enrollment, DPAN will only output predictions to any input data and not retrain on it.
\section{Conclusions and Future Work}\label{conclusion}

In this work we have proposed an approach to strongly indicate the authenticity of for noisy PUF identities based on a computer vision inspired ML classifier. By applying advanced noise tolerant classification schemes and the concept of a \textit{PUF Phenotype}, it has been shown that PUF IDs can be classified and shown to be authentic with a high degree of confidence without the need for any knowledge of the underlying PUF structure, properties, noise characteristics, or environmental conditions. This allows for robust authentication without on-device error correction and without the need for privacy leaking helper data.
The proposed approach has been verified on real DRAM Latency PUF data collected from commodity devices under a broad range of environmental conditions covering temperature and voltage fluctuation. In addition, six different classification methods were tested: XGBoost, RF, KNN, DT, LR, and SVM. Using this data classification accuracy and F1 scores of between 94\% and 98\% were achieved for varying number of supported devices respectively. We demonstrated that most tested classifiers can perform well with a tuned confidence threshold when identifying fraudulent and legitimate images, with the SVM and RF classifiers performing particularly well. It has been shown that using an appropriate confidence threshold, the risk of determining a high confidence for false responses can be eliminated while still detecting true responses on the first try in most cases. The need for occasional retries due to this is still significantly less PUF usage than would be required for high quality on-device error correction. While in this work the group membership has been treated as a static property for experimental purposes, it would be entirely possible to retain the PUF characterisation data for each device in a central and highly secure environment and use this to actively update group membership. This approach would require retraining of the model on the new set of group members, then pushing the new model out to all devices. A full exploration of this kind of approach to PUF at the protocol level may form the basis of future work.

\bibliographystyle{IEEEtran}
\bibliography{custom_bib}



\section*{Supplementary material}

# Supplementary Material

## Appendix A
### Full VGG16 Feature Extractor Confusion Matrices

We provide the confusion matrices for the test performance of each classifier when using the full VGG16 feature extractor. Figures A.1 and A.2 show the actual class and the class predicted by the model for each classifier for each test sample. Each includes the number of samples predicted for a given class alongside the percentage of predictions of each class across all predictions. Given perfect classifier performance, a confusion matrix should show a value of zero for all cells where the actual and predicted label rows and columns match. For the classifiers using the full VGG16 feature extractor, while it can be seen most classifications are correct, there are frequently classifications that are made outside of the central set of ideal classifications. For the DT classifier for example, many of each class were predicted incorrectly, showing green in many of the cells where actual and predicted label do not match. Overall, there is no significant apparent correlation to be drawn between the predicted value of certain classes and the actual value as incorrect classifications are mostly arbitrarily distributed around the confusion matrices. In some cases, there are similarities however, for example with the RF, SVM and XGB models, the a three images from device *epsilon* were classified as one of *alpha*, *beta* and *delta* respectively. This result is likely caused by the classifiers drawing similar conclusions from the features extracted from identical images impacted by in-image noise.

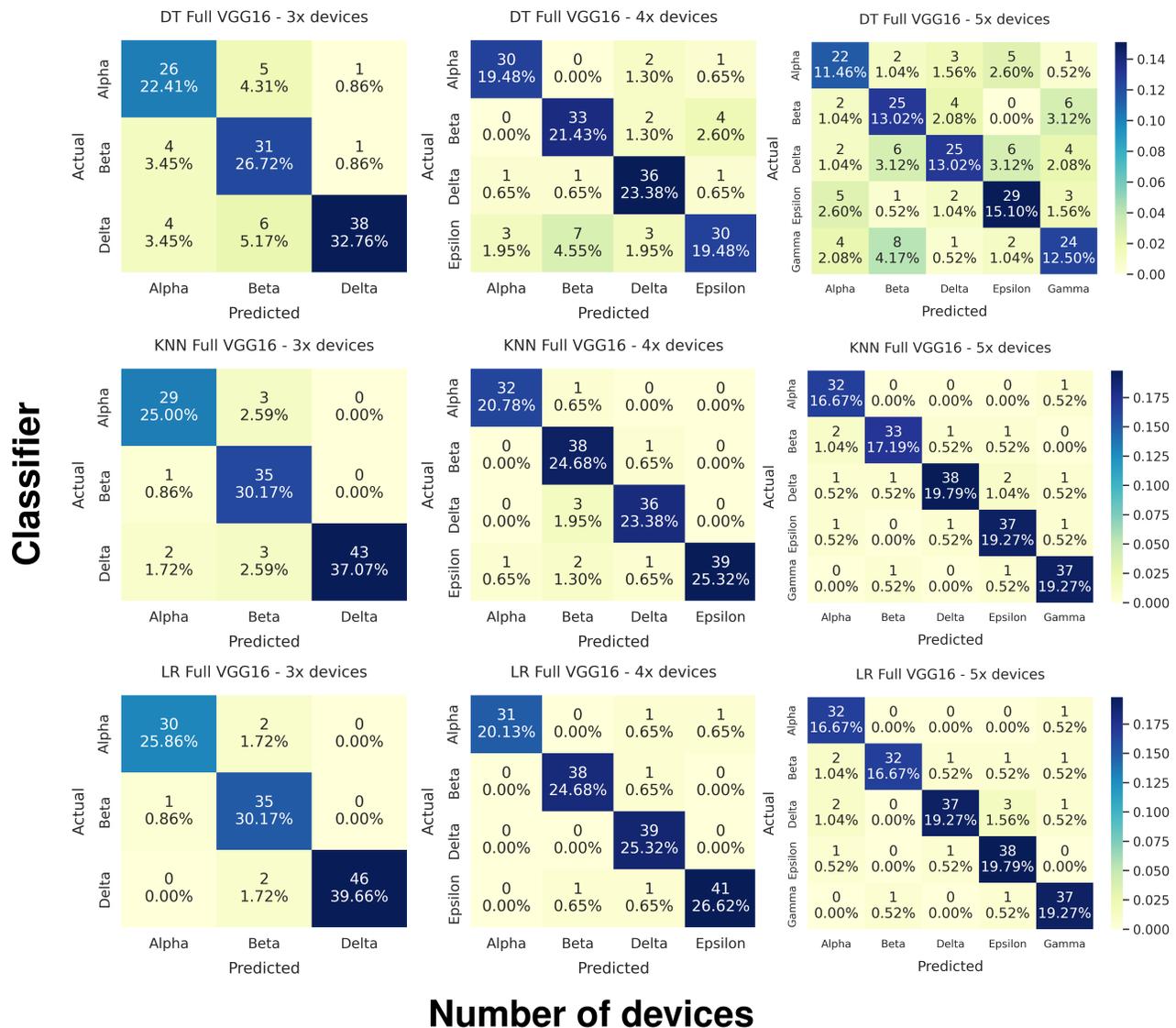

Fig. A.1. Confusion Matrices for DT, KNN & LR Classifiers Using Full VGG16 Feature Extractor

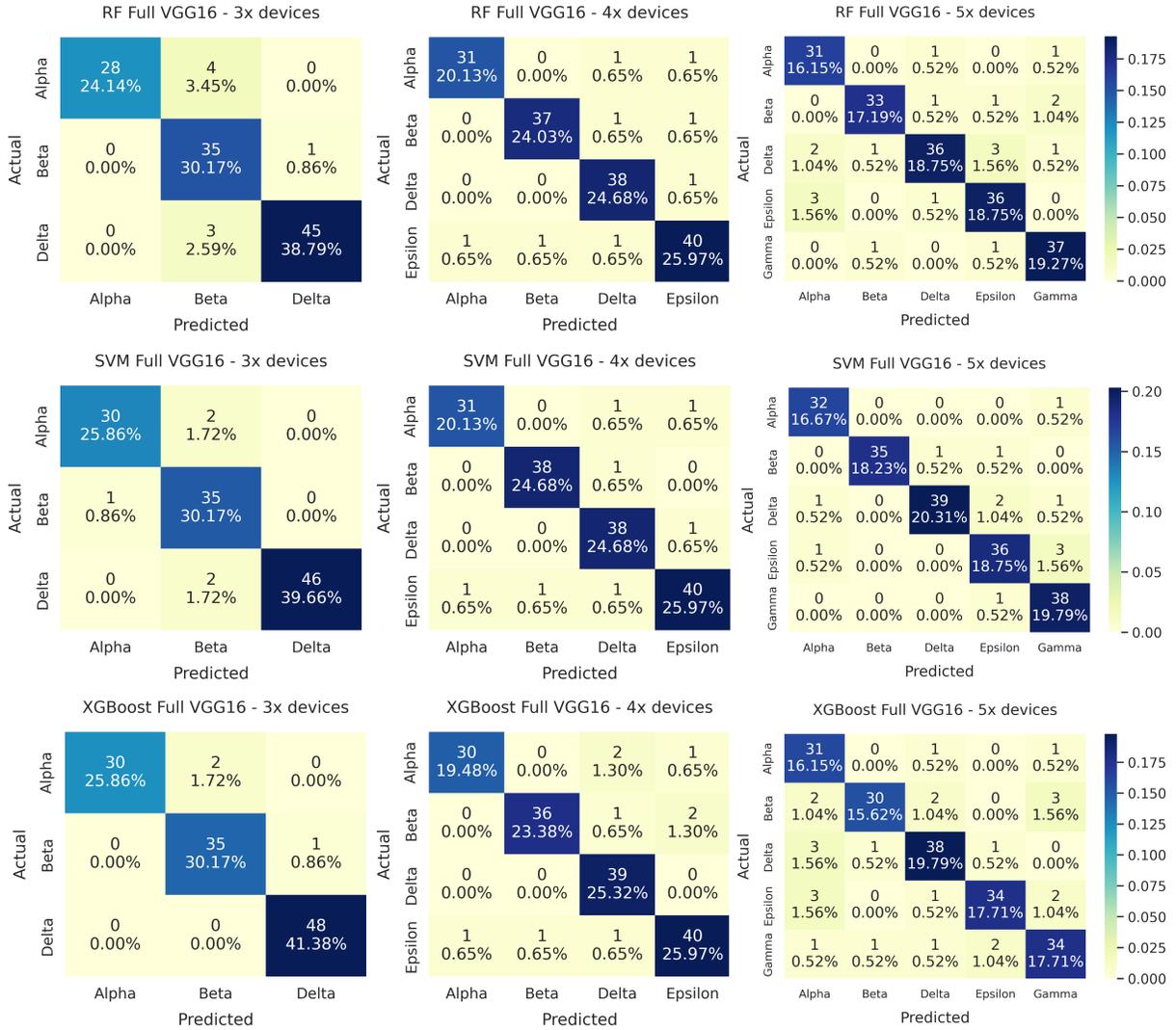

Fig. A.2. Confusion Matrices for RF, SVM & XGB Classifiers Using Full VGG16 Feature Extractor

APPENDIX B

LIGHTWEIGHT VGG16 FEATURE EXTRACTOR CONFUSION MATRICES

Here we provide the confusion matrices for each classifier when using the modified lightweight VGG16 feature extractor (excluding RF and SVM as these are provided in Figure 7 of Section IV of the main paper). This lightweight feature extractor has the dense layers of the VGG16 removed, and is the basis of our proposed scheme. As can be seen in each confusion matrix, most of the classifiers performed very strongly in their predictions for each test sample, with only few samples displayed outside of the centre line.

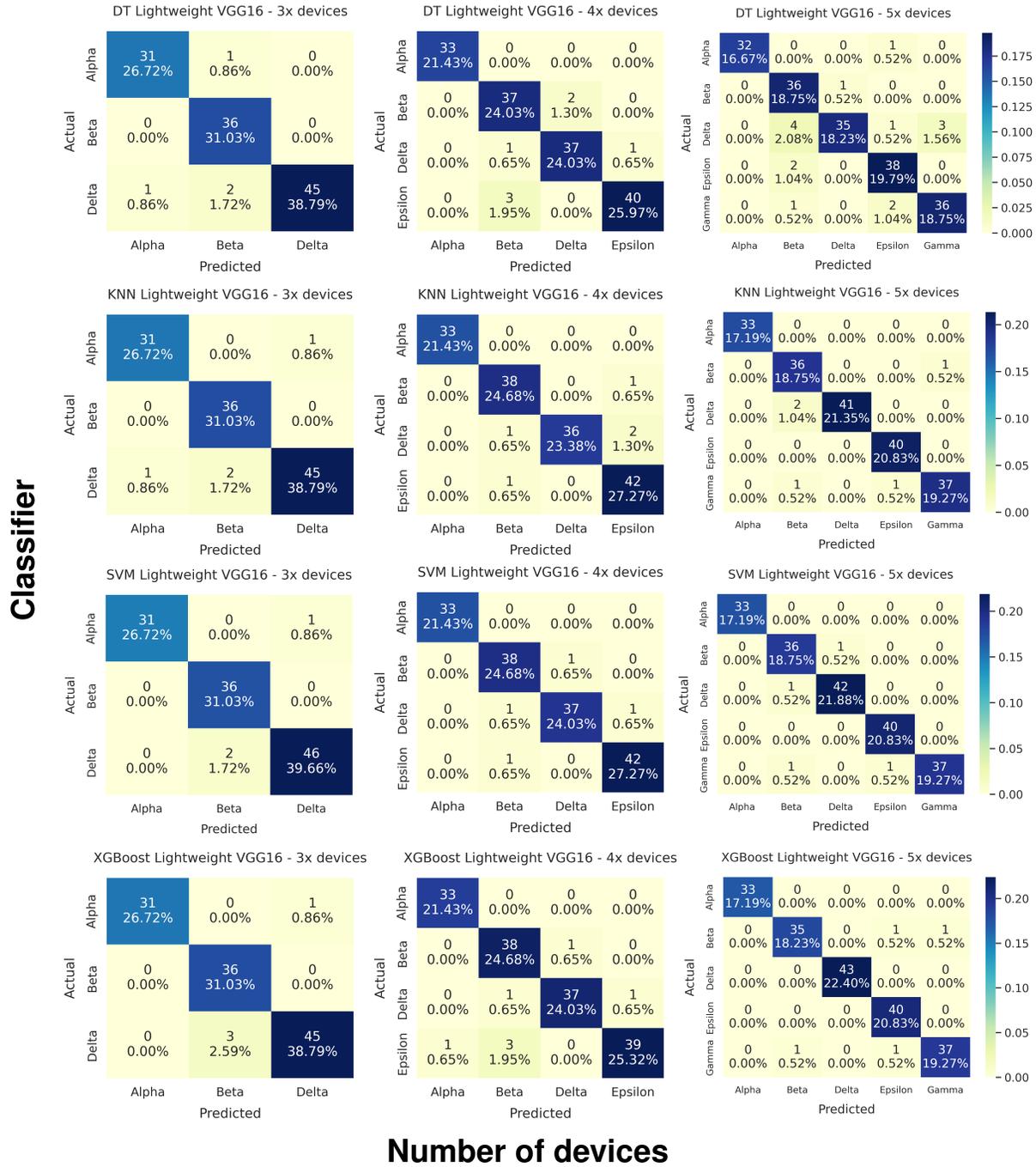

Fig. B.3. Confusion Matrices for DT, KNN, SVM & XGB Classifiers Using Lightweight VGG16 Feature Extractor

## Appendix C
### Evaluation Metrics

*Accuracy:* This metric measures the rate of correct classifications (true positive and true negative) relative to the overall number of test samples. Commonly, pure accuracy is a poor metric for classifiers trained on unbalanced data, for example in intrusion detection, where some samples are positive (intrusion detected) and most are negative (no intrusion detected), causing a potential bias. For our proposed system, we do not consider negative samples as the dataset is mostly balanced (number of samples per device) and as any classification is a positive result, the model can only either be correct or incorrect. Accuracy is defined as:

$$Accuracy = \frac{Correct\ Predictions}{Total\ Predictions} \tag{1}$$

*F1 Score:* As there is a very small variation in the number of samples for each device (class), we also consider F1 score as an evaluation metric. This is because while we do not need to consider negative results, F1 score remains an appropriate indication of performance where data is very slightly unbalanced. The F1 score is the harmonic mean of both Precision and Recall where:

$$Precision = \frac{TP}{TP + FP} \tag{2}$$

and:

$$Recall = \frac{TP}{TP + FN} \tag{3}$$

where:
- *TP* = True Positive
- *TN* = True Negative
- *FP* = False Positive
- *FN* = False Negative

F1 score is defined as:

$$F1 = \frac{2 * Precision * Recall}{Precision + Recall} = \frac{2 * TP}{2 * TP + FP + FN} \tag{4}$$

*Cross Validation:* In K-fold cross validation, the model is individually trained on each K-fold segment, with a different fold being used as the test data during each training session. The results from each test are averaged to provide an overall model performance over the full dataset. This is particularly useful in scenarios where the size of the dataset is limited as the model is able to be trained on more data. From this, we evaluate Cross-Validated Mean Accuracy and the Standard Deviation in performance across each $K$-fold.

## Appendix D
## Classifier Information

- **Decision Tree:** A Decision Tree is a simple supervised learning algorithm than can be used for regression and classification problems. As the name suggests, decision trees build tree-like structures based on rules formed given observations on features used during training. In the tree, leaves represent class labels and branches represent variations of features that lead to given class labels. Decision trees are extremely lightweight and easy to interpret as the tree can be formed visually to determine each rule based on any given input.

- **XGBoost:** XGBoost is an ensemble learning method and an implementation of gradient boosted decision trees, widely spread in applied ML. It is the most efficient and commonly used implementation of gradient boosting algorithms to date. XGBoost combines software and hardware effectively to produce superior results based on function approximation by optimizing specific loss functions and applying several regularization techniques. This has the effect of reducing the required computing resources and execution time. Key advantages include the handling of missing data automatically, parallel tree construction and continuation of training to further boost a trained model on new training data.

- **K-Nearest Neighbours:** The K-Nearest Neighbors algorithm (KNN) is a simple yet efficient classification method. The input data is classified to the most common class among its $k$ nearest neighbors when plotted ($k$ is a positive and typically small integer). The neighbors are taken from data belonging to the same class. The optimal choice of the value $k$ is highly data-dependent: in general a larger $k$ suppresses the effects of noise, but makes the classification boundaries less distinct.

- **Support Vector Machines:** Support vector machines (SVMs) are a set of supervised learning methods which can be applied to classification, regression and outlier detection problems. SVMs are one of the most robust prediction methods based on statistical learning frameworks, mapping training examples to points in space and optimising a plane based on the maximum distance between distance points of differing classes (known as the support vectors). New examples are then mapped into that same space and are labelled with a class based on which side of the plane they appear. It is also effective in high dimensional spaces and in cases where the number of dimensions is greater than the number of samples.

- **Random Forest:** Random Forest is a yet another ensemble method, more specifically bagging, which combines basic decision trees to become a robust classifier. As the random forest results are the average of the decision trees, it can significantly reduce variance.

- **Logistic Regression:** Logistic regression is a generalised linear model that can be used for classification and is also known as logit regression, maximum-entropy classification (MaxEnt) or the log-linear classifier. The probabilities representing the possible outcomes of an instance are classified via the logistic function.

APPENDIX E

MODEL HYPERPARAMETERS

For reproducibility, we provide the hyperparameters used for each classifier after tuning in Table II. We performed each experiment using the Scikit Learn library in Python. For brevity, any model hyperparameters not included in Table II were left as the default.

TABLE I

HYPERPARAMETER TUNING

| Parameter | Description | Optimal value |
|---|---|---|
| **Logistic Regression** | | |
| c | Type of regularization | 1 |
| Max Iterations | Number of iterations | 100 |
| **K-Nearest Neighbour** | | |
| n_neighbors | Number of neighbours | 9 |
| Leaf Size | Min points in node | 1 |
| **XGBoost** | | |
| ETA | Step size shrinkage | 0 |
| Gamma | Min loss reduction required to make partition on leaf node | 0 |
| Learning Rate | Rate of model learning during training | 0.02 |
| Estimators | Number of trees | 64 |
| Max Depth | Max depth of a tree | 3 |
| Min Child Weight | Min sum of instance weight needed in child | 1 |
| Subsample Ratio | Training data sample ratio prior to tree growth | 0.6 |
| **Support Vector Machine** | | |
| C | Type of regularization | 10 |
| Gamma | Kernel coefficient | 0.1 |
| **Decision Tree** | | |
| Criterion | Function used to measure the quality of split | Gini |
| Max Depth | Maximum tree depth | 10 |
| Leaf Min Samples | Min number of samples required at leaf node. | 5 |
| **Random Forest** | | |
| Estimators | Number of trees | 396 |

APPENDIX F

CONFIDENCE THRESHOLDS

We determined confidence thresholds for each classifier determined by the maximum observed confidence scores for each classifier on fraudulent input images. Each value determines the confidence output required by images to be deemed authentic by each classifier (lower is better).

TABLE II
HYPERPARAMETER TUNING

| Number of Devices | Confidence Threshold |
|---|---|
| **Logistic Regression** | |
| 3x Devices | 0.96 |
| 4x Devices | 0.95 |
| 5x Devices | 0.91 |
| **K-Nearest Neighbour** | |
| 3x Devices | 0.68 |
| 4x Devices | 0.80 |
| 5x Devices | 0.79 |
| **XGBoost** | |
| 3x Devices | 0.42 |
| 4x Devices | 0.67 |
| 5x Devices | 0.65 |
| **Support Vector Machine** | |
| 3x Devices | 0.78 |
| 4x Devices | 0.40 |
| 5x Devices | 0.42 |
| **Random Forest** | |
| 3x Devices | 0.83 |
| 4x Devices | 0.65 |
| 5x Devices | 0.68 |


\end{document}